\documentclass[11pt]{article}
\usepackage{axodraw}
\usepackage{epsfig}
\usepackage{amsfonts}
\usepackage{amsmath}
\usepackage{bm,bbm}
\usepackage{cite}
\allowdisplaybreaks[1]

\input pix.sty

\newcommand{\newv}{\mathrm{v}}
\newcommand{\lnf}{l^{ }_\rmi{1f}}
\newcommand{\lif}{l^{ }_\rmi{2f}}
\newcommand{\lnb}{l^{ }_\rmi{1b}}
\newcommand{\lib}{l^{ }_\rmi{2b}}

\newcommand{\I}{\rmii{$I$}}
\newcommand{\J}{\rmii{$J$}}
\newcommand{\sL}{\rmii{$L$}}

\newcommand{\muB}{\mu_\rmii{$B$}}
\newcommand{\tmuBL}{\tilde{\mu}_\rmii{$B$+$L$}}
\newcommand{\muY}{\mu_\rmii{$Y$}}
\newcommand{\bmuY}{\bar{\mu}_\rmii{$Y$}}
\newcommand{\bmuB}{\bar{\mu}_\rmii{$B$}}
\newcommand{\muH}{\mu_\rmii{$H$}}
\newcommand{\muLa}{\mu_{\rmii{$L$}a}}
\newcommand{\mutL}{\mu_{t_\rmii{L}}}
\newcommand{\mutR}{\mu_{t_\rmii{R}}}

\newcommand{\nG}{n_\rmii{$G$}}
\newcommand{\mW}{m_\rmii{$W$}}

\newcommand{\mH}{m_\rmii{$H$}}

\newcommand{\ko}{k_0}

\newcommand{\qm}{q_-}
\newcommand{\qp}{q_+}
\newcommand{\aL}{a^{ }_\rmii{L}}
\newcommand{\aR}{a^{ }_\rmii{R}}
\renewcommand{\eq}{eq.~}
\renewcommand{\eqs}{eqs.~}
\renewcommand{\se}{sec.~}
\renewcommand{\ses}{secs.~}
\renewcommand{\fig}{fig.~}
\renewcommand{\figs}{figs.~}

\newcommand{\alphas}{\alpha_{\rm s}}

\newcommand{\Nc}{N_{\rm c}}

\newcommand{\rmO}{{\mathcal{O}}}

\def\lsi{\raise0.3ex\hbox{$<$\kern-0.75em\raise-1.1ex\hbox{$\sim$}}}
\def\gsi{\raise0.3ex\hbox{$>$\kern-0.75em\raise-1.1ex\hbox{$\sim$}}}
\newcommand{\lsim}{\mathop{\lsi}}
\newcommand{\gsim}{\mathop{\gsi}}

\newcommand{\nF}{n_\rmii{F}}
\newcommand{\nB}{n_\rmii{B}}

\newcommand{\rmii}[1]{{\mbox{\tiny\rm{#1}}}}

\newcommand{\re}{\mathop{\mbox{Re}}}
\newcommand{\im}{\mathop{\mbox{Im}}}

\newcommand{\Tint}[1]{{\hbox{$\sum$}\!\!\!\!\!\!\!\int\,}_{\!\!\!\!\raise-0.9ex\hbox{$\scriptstyle{#1}$}}}
\newcommand{\Tinti}[1]{{{\Sigma}\!\!\!\!\raise0.3ex\hbox{$\int$}_\rmii{${#1}$}}}

\newcommand{\unit}{{\mathbbm{1}}} 
\newcommand{\bi}{\begin{itemize}}
\newcommand{\ei}{\end{itemize}}
\newcommand{\hide}[1]{ }
\newcommand{\bsl}[1]{\,\slash\!\!\!\!{#1}\,}
\newcommand{\msl}[1]{\,\slash\!\!\!{#1}\,}
\def\TAsc(#1,#2)(#3,#4,#5)%
{\SetWidth{2.0}\CArc(#1,#2)(#3,#4,#5)\SetWidth{1.0}}
\def\Lwidth{3}

\def\TAgl(#1,#2)(#3,#4,#5){\SetWidth{2.0}\PhotonArc(#1,#2)(#3,#4,#5){\Lwidth}%
{6.283 #3 mul 360 div #4 #5 sub #4 #5 sub mul sqrt mul Tdensity mul}%
\SetWidth{1.0}}
\def\TLgl(#1,#2)(#3,#4){\SetWidth{2.0}\Photon(#1,#2)(#3,#4){\Lwidth}
{#1 #3 sub #1 #3 sub mul #2 #4 sub #2 #4 sub mul add sqrt Tdensity mul}%
\SetWidth{1.0}}

\renewcommand{\picc}[1]{\;\parbox[c]{60pt}{\begin{picture}(60,30)(0,0)
\SetWidth{1.0}\SetScale{1.3} #1 \end{picture}}\; }
\def\Lwidth{1.3}

\renewcommand{\pic}[1]{\;\parbox[c]{30pt}{\begin{picture}(30,30)(0,-3)
\SetWidth{1.0}\SetScale{0.8} #1 \end{picture}}\;}
\renewcommand{\picb}[1]{\;\parbox[c]{45pt}{\begin{picture}(45,30)(0,-3)
\SetWidth{1.0}\SetScale{0.8} #1 \end{picture}}\;}
\def\procG{\picb{%
 \Lqu(0,30)(15,15)%
 \Lgl(0,0)(15,15)%
 \Lqu(15,15)(29,15)%
 \Lsc(31,14)(45,0)%
 \Line(29,15)(44,30)%
 \Line(31,14)(45,28)%
}}
\def\procGa{\picb{%
 \Lsc(0,30)(15,15)%
 \Lgl(0,0)(15,15)%
 \Lsc(15,15)(29,15)%
 \Laqu(31,14)(45,0)%
 \Line(29,15)(44,30)%
 \Line(31,14)(45,28)%
}}
\def\procH{\pic{%
 \Lqu(0,30)(15,26)%
 \Lgl(0,0)(15,4)%
 \Lsc(15,4)(15,24)%
 \Lsc(15,4)(30,0)%
 \Line(15,24)(30,28)%
 \Line(14,26)(30,30)%
}}
\def\procHa{\pic{%
 \Lsc(0,30)(15,26)%
 \Lgl(0,0)(15,4)%
 \Lqu(15,4)(15,24)%
 \Laqu(15,4)(30,0)%
 \Line(15,24)(30,28)%
 \Line(14,26)(30,30)%
}}
\def\procI{\picb{%
 \Lqu(0,30)(12.5,17)%
 \Line(12.5,17)(25,4)%
 \Lsc(0,0)(25,26)%
 \Lqu(25,4)(25,24)%
 \Lgl(25,4)(40,0)%
 \Line(25,24)(40,28)%
 \Line(24,26)(40,30)%
}}
\def\procJ{\pic{%
 \Lqu(0,30)(15,26)%
 \Lsc(0,0)(15,4)%
 \Lsc(15,4)(15,24)%
 \Lgl(15,4)(30,0)%
 \Line(15,24)(30,28)%
 \Line(14,26)(30,30)%
}}
\def\procKa{\picb{%
 \Lqu(0,30)(15,15)%
 \Laqu(0,0)(15,15)%
 \Lsc(15,15)(29,15)%
 \Laqu(31,14)(45,0)%
 \Line(29,15)(44,30)%
 \Line(31,14)(45,28)%
}}
\def\procKb{\pic{%
 \Lqu(0,30)(15,26)%
 \Lqu(0,0)(15,4)%
 \Lsc(15,4)(15,24)%
 \Lqu(15,4)(30,0)%
 \Line(15,24)(30,28)%
 \Line(14,26)(30,30)%
}}
\def\procKc{\pic{%
 \Lqu(0,30)(15,26)%
 \Laqu(0,0)(15,4)%
 \Lsc(15,4)(15,24)%
 \Laqu(15,4)(30,0)%
 \Line(15,24)(30,28)%
 \Line(14,26)(30,30)%
}}
\def\xproc{\pic{%
 \Laqu(26,14)(50,0)%
 \Lsc(0,15)(24,15)%
 \Line(24,15)(49,30)%
 \Line(26,14)(50,28)%
}}
\def\xprocG{\pic{%
 \Lgl(13,5)(13,15)%
 \Lgl(0,1)(13,5)%
 \Lgl(13,5)(26,1)%
 \Laqu(26,14)(50,0)%
 \Lsc(0,15)(24,15)%
 \Line(24,15)(49,30)%
 \Line(26,14)(50,28)%
}}
\def\xprocH{\pic{%
 \Lgl(33,-1.6)(38,7)%
 \Lgl(20,2)(33,-1.6)%
 \Lgl(33,-1.6)(42,-10.6)%
 \Laqu(26,14)(38,7)%
 \Laqu(38,7)(50,0)%
 \Lsc(0,15)(24,15)%
 \Line(24,15)(49,30)%
 \Line(26,14)(50,28)%
}}
\def\yproc{\pic{%
 \Lsc(26,14)(50,0)%
 \Lqu(0,15)(24,15)%
 \Line(24,15)(49,30)%
 \Line(26,14)(50,28)%
}}
\def\yprocG{\pic{%
 \Lgl(13,5)(13,15)%
 \Lgl(0,1)(13,5)%
 \Lgl(13,5)(26,1)%
 \Lsc(26,14)(50,0)%
 \Lqu(0,15)(14,15)%
 \Lqu(14,15)(24,15)%
 \Line(24,15)(49,30)%
 \Line(26,14)(50,28)%
}}
\def\yprocH{\pic{%
 \Lgl(33,-1.6)(38,7)%
 \Lgl(20,2)(33,-1.6)%
 \Lgl(33,-1.6)(42,-10.6)%
 \Lsc(26,14)(50,0)%
 \Lqu(0,15)(24,15)%
 \Line(24,15)(49,30)%
 \Line(26,14)(50,28)%
}}
\def\zproc{\pic{%
 \Lsc(0,0)(25,14)%
 \Lqu(0,30)(25,16)%
 \Line(25,16)(50,16)%
 \Line(25,14)(50,14)%
}}
\def\zprocG{\pic{%
 \Lsc(0,0)(25,14)%
 \Lqu(0,30)(12.5,23)%
 \Lqu(12.5,23)(25,16)%
 \Line(25,16)(50,16)%
 \Line(25,14)(50,14)%
 \Lgl(12.5,23)(17.5,31.6)%
 \Lgl(17.5,31.6)(30.5,28)%
 \Lgl(8.5,40.6)(17.5,31.6)%
}}
\def\zprocH{\pic{%
 \Lsc(0,0)(25,14)%
 \Lqu(0,30)(25,16)%
 \Line(25,16)(50,16)%
 \Line(25,14)(50,14)%
 \Lgl(12.5,7)(17.5,-1.6)%
 \Lgl(17.5,-1.6)(30.5,2)%
 \Lgl(8.5,-10.6)(17.5,-1.6)%
}}

\makeatletter \@addtoreset{equation}{section} \makeatother
\renewcommand{\theequation}{\arabic{section}.\arabic{equation}}
\makeatletter
\renewcommand\section{\@startsection {section}{1}{\z@}%
                                   {-5.5ex \@plus -1ex \@minus -.2ex}
                                   {2.3ex \@plus.2ex}%
                                   {\normalfont\large\bfseries}}
\renewcommand\subsection{\@startsection{subsection}{2}{\z@}%
                                     {-3.25ex\@plus -1ex \@minus -.2ex}%
                                     {1.5ex \@plus .2ex}%
                                     {\normalfont\normalsize\bfseries}}
\renewcommand\thesection {\@arabic\c@section}
\renewcommand\thesubsection   {\thesection.\@arabic\c@subsection}
\renewcommand{\@seccntformat}[1]{%
\csname the#1\endcsname.\hspace{1.0em}}
\makeatother

\begin{document}

\flushbottom

\begin{titlepage}

\begin{flushright}
May 2017
\vspace*{1cm}
\end{flushright}
\begin{centering}
\vfill

{\Large{\bf
 GeV-scale hot sterile neutrino oscillations: \\[3mm] 
 a derivation of evolution equations
}} 

\vspace{0.8cm}

J.~Ghiglieri$^\rmi{a}$ and M.~Laine$^\rmi{b}$
 
\vspace{0.8cm}

$^\rmi{a}${\em
Theoretical Physics Department, CERN, \\ 
CH-1211 Geneva 23, Switzerland \\}

\vspace{0.3cm}

$^\rmi{b}${\em
AEC, 
Institute for Theoretical Physics, 
University of Bern, \\ 
Sidlerstrasse 5, CH-3012 Bern, Switzerland \\}

\vspace*{0.8cm}

\mbox{\bf Abstract}
 
\end{centering}

\vspace*{0.3cm}
 
\noindent
Starting from operator equations of motion and making arguments
based on a separation of time scales, a set of equations is
derived which govern the non-equilibrium time evolution of a GeV-scale
sterile neutrino density matrix and active lepton number densities at
temperatures $T\gsim 130$~GeV. The density matrix possesses 
generation and helicity indices; we demonstrate how helicity 
permits for a classification of various sources 
for leptogenesis. The coefficients parametrizing the
equations are determined to leading order in Standard Model couplings,
accounting for the LPM resummation of $1+n \leftrightarrow
2+n$ scatterings and for all $2\leftrightarrow 2$ scatterings. The
regime in which sphaleron processes gradually decouple so that baryon
plus lepton number becomes a separate non-equilibrium variable is also
considered. 

\vfill

 
\vspace*{1cm}
  

\vfill

\end{titlepage}

%
\section{Introduction}

Relating the observed matter-antimatter asymmetry
to experimentally verifiable laws of nature is a
central challenge for cosmology. An interesting window 
of opportunity is offered by
the so-called SHiP experiment, which aims to search for GeV
scale sterile neutrinos~\cite{ship}. Following an idea 
put forward by Akhmedov, Rubakov and Smirnov~\cite{ars} and 
refined by Asaka and Shaposhnikov~\cite{as}, GeV scale sterile neutrinos
might contribute to the matter-antimatter asymmetry. This framework
is referred to as leptogenesis through sterile neutrino oscillations. 

According to Sakharov, any theoretical explanation 
of the observed matter-antimatter
asymmetry needs to come with several non-trivial ingredients: CP violation, 
deviation from thermal equilibrium, and baryon number violation. Accounting
systematically for such processes in the environment of the early universe
filled with a Standard Model plasma is a daunting task. However gradual 
progress is being made 
(cf.\ e.g.\ refs.~\cite{singlet,gagnon,shintaro,dg,canetti,
shuve,hk,abada,val,teresi,n2,n3}), 
with the goal of moving from model
computations towards first principles analyses. 

In order to formulate the computation 
in a transparent way,
it is helpful to factorize the system into 
``fast'' and ``slow'' modes. The purpose of the current paper
is to derive evolution equations for the slow modes, by
``integrating out'' the fast ones which are in thermal equilibrium.
As slow variables we take lepton and baryon asymmetries, 
and a sterile neutrino density matrix which depends on 
momentum, generation, and helicity.  
We find that both helicity states play a role, and in the presence
of lepton asymmetries they are produced and equilibrate 
at different rates.\footnote{%
 The role of helicity has recently been discussed in a different
 mass and temperature range in ref.~\cite{lello}.
 The model considered involves however a Dirac
 rather than Majorana sterile neutrino, so that 
 helicity effects are qualitatively different. 
 } 
The numerical solution of the slow dynamics
within an expanding background poses a 
challenge of its own, to be tackled in future work. 

Our plan is the following. After a formal
derivation of the basic equations in \se\ref{se:derivation}, 
we present a resummed 
perturbative determination of the coefficients appearing
in these equations in \se\ref{se:coeffs}. The right-hand sides of the
equations contain lepton and baryon 
chemical potentials: the relations of these
to lepton and baryon number 
densities are recalled in \se\ref{se:mu},
accounting properly for the (hyper)charge neutrality of the plasma. 
The evolution equation
for baryon asymmetry is given in \se\ref{se:baryon}, and we
conclude with a short outlook in \se\ref{se:concl}. 

%
\section{Derivation of evolution equations}
\la{se:derivation}

%
\subsection{Formulation of a non-equilibrium problem}

We consider temperatures above $T \sim 130$~GeV, so that 
baryon plus lepton number can change by sphaleron processes 
which are fast enough to be  
in or close to thermal equilibrium~\cite{sphaleron}. The crossover at which 
the electroweak symmetry gets ``restored'' is  
at $T \sim 160$~GeV~\cite{crossover,dono}, however for GeV-scale sterile 
neutrinos the rates that we are
interested in do not change much in the temperature range between
130~GeV and 160~GeV~\cite{broken}. Therefore, for the conceptual
discussion, we can imagine to work in the ``symmetric'' phase
of the electroweak theory.\footnote{%
 For the numerical analysis, infrared (IR) sensitive effects from
 $1+n \leftrightarrow 2+n$ scatterings, described in \se\ref{ss:1to2}, 
 need however to be separately implemented for the symmetric and 
 the Higgs phases~\cite{broken}. 
 } 

In the temperature range 130~GeV $\lsim T \lsim $ 10$^5$~GeV, 
all Standard Model interactions can be 
assumed to be in thermal equilibrium
(this includes lepton chirality flipping processes
through the electron Yukawa coupling). Then 
the state of the system is characterized by a 
temperature, $T$, by three lepton chemical potentials, $\mu^{ }_a$, 
and by the baryon chemical potential, $\muB$.  Suppose now that  
we extend the Standard Model through right-handed  
sterile (gauge singlet) neutrinos, and use these to 
generate active neutrino masses through the see-saw mechanism. 
If the mass of the sterile neutrinos is $\sim$ GeV, then
the Yukawa couplings are so small ($\sim 10^{-7}$) that sterile
neutrinos do {\em not} equilibrate in this temperature range. 
Therefore they constitute a non-equilibrium ensemble, 
evolving ``slowly'' in a Standard Model background. The Yukawa
interactions also imply that lepton number densities are not conserved, 
so these become slowly evolving variables as well. 

The neutrino Yukawa interactions need not be 
aligned with Standard Model lepton generations. This leads
to sterile neutrino oscillations. Because the neutrino Yukawa
couplings are tiny, coherent oscillations may be maintained for
a long period of time, and sterile neutrinos need to be 
described by a density matrix. 

The momenta 
of the sterile neutrinos 
are changed by the same slow interactions as their
number densities are. 
Therefore, kinetic equilibrium cannot be assumed and 
the density matrix displays a non-trivial momentum dependence. 

There is one further slow variable to be tracked, 
namely the helicity of the sterile neutrinos. As massive particles, 
sterile neutrinos can carry both helicities. 
The two helicity states experience
different interactions: basically, one state interacts with 
Standard Model leptons and the other with antileptons. 
Both states need to be included in the density matrix. 

To summarize, we 
need a density matrix for each sterile neutrino momentum mode, labelled
by $k \equiv |\vec{k}|$. It turns out that to a good approximation
the two helicity states have no direct overlap with  
each other (cf.\ discussion at the beginning of \se\ref{ss:summary}). 
Therefore, for each $k$
the set of non-equilibrium variables 
consists of two complex matrices, 
denoted by $\rho^{ }_{(\tau)\I\J}$, where 
$\tau = \pm$ labels helicity and $I,J$ label generations. 
In addition the three active 
lepton asymmetries, denoted by $n^{ }_a$, 
and the baryon asymmetry, denoted by $\nB$, evolve slowly.  

The goal of our study is to derive evolution equations for the 
non-equilibrium variables to $\rmO(h^2)$ in neutrino 
Yukawa couplings. In principal the general form of the equations 
is valid to all orders in Standard Model couplings (in practice
certain small corrections are omitted along the way), however we subsequently
evaluate the coefficients at leading order (cf.\ \se\ref{se:coeffs}).
The physically interesting effects, 
which are of $\rmO(h^4)$ or $\rmO(h^6)$~\cite{shuve}, originate 
from the coupled dynamics of the ``slow'' oscillations, and will
be addressed in a separate study. 

%
\subsection{Basic variables and equations of motion}

In the temperature range considered ($T\gsim 130$~GeV) the Higgs mechanism 
gives a contribution small compared with thermal masses, 
i.e.\ $\mW \sim gv/2 \ll g T$, where $g$ denotes the SU(2) gauge coupling. 
Then the masses of sterile neutrinos are directly
given by the Majorana masses, assumed real and positive and denoted
by $M^{ }_\I$, $I \in \{1,2,3\}$. The Lagrangian reads 
\be
 \mathcal{L} = \mathcal{L}^{ }_\rmii{SM} 
 + \fr12 \sum_{\I} \bar{N}^{ }_\I 
 \bigl( i \gamma^\mu\partial_\mu - M^{ }_\I \bigr) {N}^{ }_\I
 - \sum_{\I,a}
 \Bigl(
  \bar{\ell}^{ }_{a}  \aR \tilde{\phi}\, h^{*}_{\I a}\, N^{ }_\I
  +
  \bar{N}^{ }_\I \, h^{ }_{\I a}\, \tilde{\phi}^\dagger  \aL \ell^{ }_a
 \Bigr)
 \;, \la{L_Majorana}
\ee
where $\tilde{\phi} = i \sigma_2 \phi^*$ 
is a Higgs doublet; $\aL, \aR$ are chiral projectors; 
$\ell^{ }_a = (\nu\, e)^T_a$ is a left-handed lepton doublet
of generation $a$; and $h^{ }_{\I a}$ are the components
of the neutrino Yukawa matrix. 

We consider the so-called ultrarelativistic regime, 
$k\sim \pi T \gg M^{ }_\I$, so that a free dispersion relation reads 
\be
 \omega^k_\I
 \; \equiv \; \sqrt{k^2 + M_\I^2}
 \; \approx \; k + \frac{M_\I^2}{2k}
 \;. \la{ur}
\ee 
In this regime the vacuum mass is corrected 
by a thermal effect of $\rmO(h^2T^2)$~\cite{weldon}.
Even though $h$ is small, the thermal correction is  
relevant because it should be compared with the mass differences 
$M_\I^2 - M_\J^2$ and because the initial temperature
may be high, $T \sim 10^5$~GeV. In our formalism, thermal
masses originate as a part of the $\rmO(h^2)$ corrections, 
cf.\ appendix~A.  
Therefore, we treat the kinematics as in vacuum for the moment. 
The kinematic approximation of \eq\nr{ur} is frequently
invoked in order to simplify the discussion. 

The sterile neutrino 
field operator in the interaction picture can be written as 
\be
 N^{ }_\I(\mathcal{X}) = 
 \int_{\vec{k}} \frac{1}{\sqrt{2 \omega_\I^k}} 
 \sum_\tau 
 \Bigl( 
   u^{ }_{\vec{k}\tau\I} \,
   a^{ }_{\vec{k}\tau\I} \,
   e^{- i \mathcal{K}^{ }_\I \cdot \mathcal{X}}
  +  
   \newv^{ }_{\vec{k}\tau\I} \,
   a^{\dagger}_{\vec{k}\tau\I} \,
   e^{i \mathcal{K}^{ }_\I \cdot \mathcal{X}}
 \Bigr)
 \;, \la{onshell}
\ee
where
$
 \int_{\vec{k}} \equiv \int \! \frac{{\rm d}^3\vec{k}}{(2\pi)^3}
$ 
and
$ 
 \mathcal{K}^{ }_\I \cdot \mathcal{X}
 \equiv
 \omega^k_\I t - \vec{k}\cdot\vec{x}
$.
In accordance with 
the Majorana nature of $N^{ }_\I$, the on-shell spinors are related
by $ \newv^{ }_{ }  =  C  \bar{u}^{T}_{ } $, 
where $ C $ is the charge conjugation matrix. 
The creation and annihilation operators, which are time-independent
in \eq\nr{onshell},  satisfy the commutation relations
$
 \{ 
    a^{ }_{\vec{k}\tau\I} , 
    a^{ \dagger }_{\vec{q}\sigma\J} 
 \}  
 =
  (2\pi)^3 \delta^{(3)}(\vec{k} - \vec{q})
  \delta^{ }_{\tau\sigma} \delta^{ }_{\I\J}
$. 

Sterile neutrinos interact through the Yukawa terms
in \eq\nr{L_Majorana}. We rephrase the interactions through
an interaction Hamiltonian,  
\be
 H^{ }_\rmi{int}(t) 
 = \int_{\vec{x}} \sum_{\I , a}
 \bigl[
   \bar{j}^{ }_a(\mathcal{X})\,  h^{*}_{\I a}\, N^{ }_{\I}(\mathcal{X})
  \; + \; 
   \bar{N}^{ }_\I(\mathcal{X})\, h^{ }_{\I a}\,  j^{ }_a(\mathcal{X})   
 \bigr]
 \;, \quad
 \mathcal{X} = (t,\vec{x})
 \;. \la{Hint}
\ee
By $j^{ }_a$ and $\bar{j}^{ }_a$ 
we denote Standard Model currents from \eq\nr{L_Majorana}, 
\be
 j^{ }_a \;\equiv\; \aL j^{ }_a \;\equiv\; \tilde{\phi}^\dagger \aL \ell^{ }_a 
 \;, \quad
 \bar{j}^{ }_a \;\equiv\; \bar{\ell}^{ }_{a}  \aR \tilde{\phi}
 \;. \la{ja} 
\ee

In order to understand the dynamics induced by $H^{ }_\rmi{int}$, 
it is helpful to go over to the Heisenberg picture
for a moment (cf.\ ref.~\cite{equil} for an analogous discussion, 
and appendix~\ref{ss:qm} for a detailed step-by-step argument). 
Then the canonical equation of motion for the annihilation 
operator, defined by expressing the field operator in the form of 
\eq\nr{onshell} and accounting for any additional time dependences 
through $a^{ }_{\vec{k}\tau\I}$ and ${a}^{ \dagger }_{\vec{q}\sigma\J}$,  
becomes
\ba
 i \dot{a}^{ }_{\vec{k}\tau\I}(t)
 & = & 
 \bigl[ a^{ }_{\vec{k}\tau\I} ,  H^{ }_\rmi{int}  \bigr] 
 \nn 
 & = & 
 \frac{1}{\sqrt{2 \omega^k_\I}}
 \int_{\vec{x}} \sum_a 
 \bigl[ 
 \bar{u}^{ }_{\vec{k}\tau\I} h^{ }_{\I a}\, j^{ }_a (\mathcal{X}) - 
 \bar{j}^{ }_a (\mathcal{X}) \,  h^*_{\I a} \newv^{ }_{\vec{k} \tau\I} 
 \bigr]
 \, e^{i \mathcal{K}^{ }_\I \cdot \mathcal{X}}
 \;. 
 \la{eom_a}
\ea
An analogous equation is obtained for
$
 {a}^{ \dagger }_{\vec{q}\sigma\J}
$. 
The canonical anticommutator remains time-independent. 
Similarly, the lepton asymmetries evolve as~\cite{washout}
\be
 i \dot{L}^{ }_a(t) 
 \; = \; 
 \int_\vec{x} \sum_\I \bigl[ 
 \bar{j}^{ }_a(\mathcal{X}) h^{*}_{\I a}\, N_\I^{ } (\mathcal{X}) - 
 \bar{N}^{ }_\I(\mathcal{X}) h^{ }_{\I a} j^{ }_a(\mathcal{X}) 
 \bigr]
 \;. \la{eom_L}
\ee 
 
For the physical observables that we are interested in, 
the evolution rate is of $\rmO(h^2)$. We extract the rate from
an expectation value of an operator like in \eq\nr{eom_L}. 
In order to evaluate the expection value, we return to the 
interaction picture. Then, the time evolution of the density
matrix is determined by $H^{ }_\rmi{int}$. In particular, 
assuming that the full density matrix is known at some 
time $t=0$, its time evolution  is to first order in $h$ given by 
\be
 \rho^{ }_\rmi{full}(t) = 
 \rho^{ }_\rmi{full}(0)  - i \int_0^t \! {\rm d}t' \,  
 \bigl[ H^{ }_\rmi{int} (t') , \rho^{ }_\rmi{full}(0) \bigr] + \rmO(h^2) 
 \;. \la{density_matrix}
\ee
The physical rate can then be 
defined as (cf.\ e.g.\ ref.~\cite{lsb}, and 
\eq\nr{detailed} for an explanation of intermediate steps)
\ba
 \langle \dot{O}(t) \rangle 
 & \equiv & 
 \tr \bigl[ \dot{O}(t) \rho^{ }_\rmi{full}(t) \bigr]
 \nn 
 & = &   
 \tr 
 \bigl\{
 \bigl[ \dot{O}(t)^{ } , 
  - i \int_0^t \! {\rm d}t' \, H^{ }_\rmi{int} (t')
 \bigr] \, \rho^{ }_\rmi{full}(0) 
 \bigr\}
 \; + \; \rmO(h^3)  
 \;. \la{eom_general}
\ea
The expectation value with respect to the density 
matrix $\rho^{ }_\rmi{full}(0)$ is denoted by 
$\langle ... \rangle \equiv \tr \{ ...\, \rho^{ }_\rmi{full}(0) \}$. 
At the end of the computation, this
can be re-interpreted as having been evaluated
with the density matrix at time $t$, since the difference between
$\rho^{ }_\rmi{full}(t)$ and $\rho^{ }_\rmi{full}(0)$ 
is of $\rmO(h)$. This way, so-called
secular terms can be avoided. 

Because of the 
different times scales related to the ``slow'' and ``fast'' 
variables, we can assume the full density matrix to have  
a block-diagonal form, 
$
 \rho^{ }_\rmi{full} = 
 \rho^{ }_\rmii{$N$} \otimes \rho^{ }_\rmii{SM}
$, 
where $\rho^{ }_\rmii{$N$}$ 
is the density matrix associated with the sterile neutrinos. 
The density matrix associated with the 
Standard Model degrees of freedom,  
$
 \rho^{ }_\rmii{SM}
$, 
is in equilibrium at a temperature $T$ and is parametrized by
(slowly evolving) chemical potentials $\mu^{ }_a$ and $\muB$: 
\be
 \rho^{ }_\rmii{SM} = \frac{1}{Z^{ }_\rmii{SM}}
 \exp\biggl(-\frac{H^{ }_\rmii{SM} - \sum_a \mu^{ }_a L^{ }_a - \muB B}{T}
 \biggr)
 \;. \la{rho_SM}
\ee
In the canonical formalism there are no other chemical potentials, 
however in the path integral formalism the hypercharge gauge field
gets an expectation value in the presence of $\mu^{ }_a,\muB \neq 0$
which effectively generates an additional chemical potential for 
all fields coupling to the hypercharge field (cf.\ \se\ref{se:mu}).

We note that the operator equation of motion
in \eq\nr{eom_L} has the form
\be
 \dot{L}^{ }_a(t) = i \int_\vec{x}
 \sum_\I \bigl[ \mathcal{H}^{ }_{\I a}(\mathcal{X}) - 
 \mathcal{H}^\dagger_{\I a} (\mathcal{X}) \bigr]
 \;, 
\ee
whereas the interaction Hamiltonian in \eq\nr{Hint} can be written as 
\be
 H^{ }_\rmi{int}(t') = \int_\vec{y}
 \sum_{\J,b} \bigl[ \mathcal{H}^{ }_{\J b}(\mathcal{Y}) + 
 \mathcal{H}^\dagger_{\J b} (\mathcal{Y})
 \bigr]
 \;, 
\ee
where $\mathcal{Y} \equiv (t',\vec{y})$. Inserting these structures into
\eq\nr{eom_general} we get
\ba
  \langle \dot{L}^{ }_a(t) \rangle & = & 
  \int_0^t \! {\rm d}t' \! \int_{\vec{x,y}} \sum_{\I,\J,b}
  \Bigl\langle
  \bigl[ 
  \mathcal{H}^{ }_{\I a}(\mathcal{X}) - 
  \mathcal{H}^\dagger_{\I a}  (\mathcal{X}) \, , \, 
  \mathcal{H}^{ }_{\J b}(\mathcal{Y}) + 
  \mathcal{H}^\dagger_{\J b} (\mathcal{Y})
  \bigr] 
  \Bigr\rangle
  + \rmO(h^3)  
  \;. \hspace*{6mm} \la{gen_1}
\ea
Here the correlators 
$
 \langle [\mathcal{H}^{ }_{\I a}(\mathcal{X}), 
 \mathcal{H}^{ }_{\J b}(\mathcal{Y})]\rangle
$
and 
$
 \langle [\mathcal{H}^\dagger_{\I a}(\mathcal{X}),
  \mathcal{H}^\dagger_{ \J b}(\mathcal{Y})]\rangle
$
vanish, because within the Standard Model there are to $\rmO(h^0)$
direct correlations
only between $j^{ }_a$ and $\bar{j}^{ }_a$. 
For the same reason, the sum
over $b$ is saturated by $b = a$. 
Furthermore, we can take the limit $t\to \infty$, given
that reaction rates proportional to $\sim h^2$
are much slower (with a much larger time scale) 
than the fast Standard Model rates. 

%
\subsection{Time evolution of a sterile neutrino density matrix}
\la{ss:rho}

Let us apply the formalism above to the time evolution
of a sterile neutrino ``density matrix''. We define it as 
\be
 \hat{\rho}^{ }_{\tau \I;\sigma\J} 
 \; \equiv \; 
 \frac{a^\dagger_{\vec{k}\tau \I} 
       a^{ }_{\vec{k}\sigma\J}}{V}
 \;,  \la{def_rho}
\ee
where $V$ is the volume, taken to be infinite at the end
of the computation. 
This operator now plays the role of $O(t)$ in \eq\nr{eom_general}. 

Making use of the equation of motion in \eq\nr{eom_a}, 
\eq\nr{eom_general} leads to 2-point correlators of the Standard Model
currents $j^{ }_a$ and $\bar{j}^{ }_a$, 
defined in \eq\nr{ja}. These can be expressed in terms
of Wightman functions,
\ba
 \bigl\langle
   j^{ }_{a \alpha }(\mathcal{X}) \, 
   \bar{j}^{ }_{b \beta }(\mathcal{Y})
 \bigr\rangle 
 & = & 
 \delta^{ }_{ab} \int_\mathcal{P} 
  e^{- i \mathcal{P}\cdot(\mathcal{X-Y})} \, 
 \Pi^{>}_{a\alpha\beta}(\mathcal{P})
 \;, \\ 
 \bigl\langle
   \bar{j}^{ }_{b \beta }(\mathcal{Y}) \, 
   j^{ }_{a \alpha }(\mathcal{X}) 
 \bigr\rangle 
 & = & 
 - \delta^{ }_{ab} \int_\mathcal{P} 
  e^{- i \mathcal{P}\cdot(\mathcal{X-Y})} \, 
 \Pi^{<}_{a\alpha\beta}(\mathcal{P})
 \;, 
\ea  
where $\alpha,\beta \in \{1,...,4\}$ represent spinor indices. 
The operators $j^{ }_a,\bar{j}_a$ have a non-trivial commutator with 
the lepton number operator $L^{ }_a$ appearing
in the density matrix $\rho^{ }_\rmii{SM}$, cf.\ \eq\nr{rho_SM}; 
as a consequence, following a text-book derivation, 
the Wightman functions can be expressed in terms of the spectral function, 
with relations 
depending on the index $a$ through the chemical
potential that is carried by active leptons: 
\ba
 \Pi^{>}_{a\alpha\beta}(\mathcal{P}) & = & 
 2 \bigl[ 1 - \nF(\omega - \mu^{ }_a) \bigr] \, 
 \rho^{ }_{a\alpha\beta}(\mathcal{P}) 
 \;, \la{rel1} \\  
 \Pi^{<}_{a\alpha\beta}(\mathcal{P}) & = & 
 - 2 \nF(\omega - \mu^{ }_a) \, 
 \rho^{ }_{a\alpha\beta}(\mathcal{P}) 
 \;, \\ 
 \Pi^{>}_{a\alpha\beta}(-\mathcal{P}) & = & 
 2 \nF(\omega + \mu^{ }_a) \, 
 \rho^{ }_{a\alpha\beta}(-\mathcal{P}) 
 \;, \\
 \Pi^{<}_{a\alpha\beta}(-\mathcal{P}) & = & 
 - 2 \bigl[ 1 - \nF(\omega + \mu^{ }_a) \bigr] \, 
 \rho^{ }_{a\alpha\beta}(-\mathcal{P}) 
 \;. \la{rel4}
\ea
Here $\mathcal{P}=(\omega,\vec{p})$ and 
$\nF(\omega) \equiv 1 / (e^{\omega / T} + 1)$ is the Fermi distribution.

In the expectation value following from \eq\nr{eom_general}, 
the spectral function is bracketed by the on-shell
spinors $u$ and $\newv$. The expression can be simplified by making use of 
the fact that, with the exception of processes involving Yukawa 
couplings, chirality is preserved by Standard Model interactions at high
temperatures. Omitting higher-order contributions involving the 
Yukawas,\footnote{%
 The leading-order contributions of $\rmO(h_t^2)$, appearing as 
 part of $2 \leftrightarrow 2$ scatterings in \se\ref{ss:22}, 
 are however kept and do not negate the argument, 
 because they originate through scalar exchange. 
 } 
chiral invariance implies that the spectral function is 
proportional to the Dirac matrices $\gamma^\mu$~\cite{weldon}. Then 
we can use the properties of the charge conjugation matrix $C$ to show that
\ba
   \bar{u}^{ }_{\vec{k}\tau\I}
   \, \aL \,\rho^{ }_{a} (\mathcal{P})\, \aR \,
   \newv^{ }_{\vec{q}\sigma\J}
 & = & 
   \bar{u}^{ }_{\vec{q}\sigma\J}
   \, \aR \,\rho^{ }_{a} (\mathcal{P})\, \aL \,
   \newv^{ }_{\vec{k}\tau\I}
 \;, \\  
   \bar{\newv}^{ }_{\vec{k}\tau\I}
   \, \aL \,\rho^{ }_{a} (\mathcal{P})\, \aR \,
   \newv^{ }_{\vec{k}\sigma\J}
 & = &
   \bar{u}^{ }_{\vec{k}\sigma\J}
   \, \aR \,\rho^{ }_{a} (\mathcal{P})\, \aL \,
   u^{ }_{\vec{k}\tau\I}
 \;. 
\ea
Furthermore, 
chiral invariance implies
that amplitudes between $\bar{u}$ and $u$ conserve helicity
(helicity states are defined in \eq\nr{u}, and examples of 
non-zero matrix elements are shown in \eq\nr{trace2}), 
\be
   \bar{u}^{ }_{\vec{k}\sigma\J}
   \, \aL \,\rho^{ }_{a} \, \aR \, 
   u^{ }_{\vec{k}\tau\I}
   \; \propto \; \delta^{ }_{\sigma\tau}
   \;, \quad
   \bar{u}^{ }_{\vec{k}\sigma\J}
   \, \aR \,\rho^{ }_{a} \, \aL \, 
   u^{ }_{\vec{k}\tau\I}
   \; \propto \; \delta^{ }_{\sigma\tau}
 \;. \la{helicity}
\ee
We make use of this important simplification in the following. 

The integration over the time $t'$ in \eq\nr{eom_general}
can also be simplified. In an equation like \eq\nr{gen_1}
we are faced with integrations of the types 
\ba
 I^{ }_1 & = &  
 e^{i (\omega^k_\J - \omega^k_\I) t}
 \int_0^t \! {\rm d}t' 
 \int_{-\infty}^{\infty} \! \frac{{\rm d}\omega}{2\pi}
 e^{i(\omega - \omega^k_\J)(t-t')}
 \,\phi^{ }_1(\omega)
 \;, \\
 I^{ }_2 & = & 
 e^{- i (\omega^k_\J + \omega^k_\I) t} 
 \int_0^t \! {\rm d}t' 
 \int_{-\infty}^{\infty} \! \frac{{\rm d}\omega}{2\pi}
 e^{i(\omega + \omega^k_\J)(t-t')}
 \,\phi^{ }_2(\omega)
 \;. 
\ea
Here $I^{ }_1$ multiplies 
terms bracketed between $\bar{u}$ and $u$, and 
$I^{ }_2$ those bracketed between $\bar{u}$ and~$\newv$. 
Given that $\omega^k_\I \approx k + M_\I^2/(2 k) \sim 3 T$, 
$I^{ }_2$ contains a rapid oscillation, similar to  
the ``fast'' Standard Model variations; these rapid oscillations
will be omitted. In $I^{ }_1$
the oscillation rate is suppressed by Majorana mass differences; 
this is a ``slow'' process and needs to be kept. 

The large-$t$ value of $I^{ }_1$ can be defined as a limit: 
\ba
 \lim_{t\to\infty} \int_0^t \! {\rm d}t' \, 
 e^{i(\omega - \omega^k_\J)(t-t')} & \equiv & 
 \lim_{\epsilon\to 0^+}
 \int_0^\infty \! {\rm d}t' \, 
 e^{i(\omega - \omega^k_\J + i \epsilon) t'}
 \;  = \; 
 \frac{i}{\omega - \omega^k_\J + i \epsilon}
 \nn & = & 
 \mathcal{P}\Bigl( \frac{i}{\omega - \omega^k_\J} \Bigr) + 
 \pi \delta(\omega - \omega^k_\J) 
 \;. \la{principal}
\ea
Assuming that $\phi^{ }_1$ is slowly varying around 
$\omega \approx \omega^k_\J$, the principal value part is  
antisymmetric around this point and corresponds 
to a higher time derivative correction; it amounts
to a modification of $\omega^k_\J$ through a thermal
mass (this is shown in appendix~A). 
We postpone the inclusion of this
``dispersive'' or ``virtual'' correction for the moment, 
focussing first on ``absorptive'' or ``real'' effects. For those, we need 
$
 \re I^{ }_1 \approx \fr12  e^{i (\omega^k_\J - \omega^k_\I) t} 
 \,\phi^{ }_1(\omega^k_\J) 
$.

Inserting the time integral as well as 
\eqs\nr{rel1}--\nr{rel4} into \eq\nr{eom_general}, we find that absorptive
time evolution is parametrized
by the slowly evolving coefficients 
\ba
 \hat{\Gamma}^+_{(a\tau)\I\J}(t) & \equiv & 
 \frac{h^*_{\I a}h^{ }_{\J a}}{\sqrt{\omega^k_\I \omega^k_\J}} \,
  \bar{u}^{ }_{\vec{k}\tau\J}
 \, \aL \, \rho^{ }_a(\mathcal{K}^{ }_\J) \, \aR \, 
 u^{ }_{\vec{k}\tau\I} \, e^{i (\omega^k_\J - \omega^k_\I) t}
 \;, \la{Gamma_p} \\
 \hat{\Gamma}^-_{(a\tau)\I\J}(t) & \equiv & 
 \frac{h^{ }_{\I a}h^{*}_{\J a}}{\sqrt{\omega^k_\I \omega^k_\J}} \,
 \bar{u}^{ }_{\vec{k}\tau\J}
 \, \aR \, \rho^{ }_a( - \mathcal{K}^{ }_\J) \, \aL \,
 u^{ }_{\vec{k}\tau\I} \, e^{i (\omega^k_\J - \omega^k_\I) t}
 \;, \la{Gamma_m}
\ea
where $\mathcal{K}^{ }_\J \equiv (\omega^k_\J,\vec{k})$.
Noting that $\rho^{ }_a$ is real
(cf.\ appendix~B of ref.~\cite{recent} for a general discussion), 
the evolution equation reads
\ba
 \langle \dot{\hat{\rho}}^{ }_{\tau\I;\sigma\J } \rangle
 & = & 
 \fr12 \sum_{\sL,a}
 \biggl\{
   \hat{\Gamma}^{+}_{(a\tau)\I\sL}(t) \, 
   \Bigl[ \delta^{ }_{\tau\sigma} \delta^{ }_{\sL\J} 
          \, \nF(\omega^k_\J - \mu^{ }_a) 
          \; - \; \langle \hat{\rho}^{ }_{\tau\sL;\sigma\J} \rangle \Bigr] 
 \nn 
 & + & 
   \Bigl[ \delta^{ }_{\tau\sigma} \delta^{ }_{\I\sL} 
          \, \nF(\omega^k_\I - \mu^{ }_a) 
          \; - \; \langle \hat{\rho}^{ }_{\tau\I;\sigma\sL} \rangle \Bigr] 
   \hat{\Gamma}^{+*}_{(a\sigma)\J\sL}(t) \, 
 \nn 
 & + & 
   \hat{\Gamma}^{-}_{(a\tau)\I\sL}(t) \, 
   \Bigl[ \delta^{ }_{\tau\sigma} \delta^{ }_{\sL\J} 
          \, \nF(\omega^k_\J + \mu^{ }_a) 
          \; - \; \langle \hat{\rho}^{ }_{\tau\sL;\sigma\J} \rangle \Bigr] 
 \nn 
 & + & 
   \Bigl[ \delta^{ }_{\tau\sigma} \delta^{ }_{\I\sL} 
          \, \nF(\omega^k_\I + \mu^{ }_a) 
          \; - \; \langle \hat{\rho}^{ }_{\tau\I;\sigma\sL} \rangle \Bigr] 
   \hat{\Gamma}^{-*}_{(a\sigma)\J\sL}(t) \, 
 \biggr\} + \rmO(h^3)
 \;, \la{evol_rho}
\ea 
where the ``equilibrium'' terms containing $\nF$
originate from  
$\{ a^{ }_{\vec{k}\sigma\J}, a^{\dagger}_{\vec{k}\tau\sL}\}/V
 = \delta^{ }_{\tau\sigma}\delta^{ }_{\sL\J}$. 
The terms containing $\hat{\Gamma}^+$ and $\nF(\omega - \mu^{ }_a)$
represent scatterings involving leptons, whereas those 
with $\hat{\Gamma}^-$ and $\nF(\omega + \mu^{ }_a)$ represent the 
contributions of antileptons. 
Physically, $ \hat{\Gamma}^+_{(a\tau)\I\J} $ describes the rate
at which the in-medium wave function of the state $(\vec{k}\tau J)$
gets projected in the direction of $(\vec{k}\tau I)$, and 
$ \hat{\Gamma}^-_{(a\tau)\I\J} $ does the same for the 
charge-conjugated process.  

It may be noted that 
the right-hand side of \eq\nr{evol_rho}
vanishes in equilibrium, i.e.\ 
if the density matrix is diagonal and
all lepton chemical potentials vanish. 
Its general form is, however,  
valid both near and far from equilibrium.

It can also be observed that
there is no equilibrium term with $\tau\neq \sigma$: 
helicity non-diagonal correlations 
decrease to zero with time. In particular, if we start from an initial 
density matrix
which is helicity-diagonal, it stays so. However, the values of
the coefficients $\hat{\Gamma}^{\pm}$ do depend on the helicity
index $\tau$ (cf.\ \se\ref{se:coeffs}). 

For formal considerations, it is convenient to have
coefficients which are independent of time. This also offers
for a simple way to include the dispersive 
thermal mass corrections mentioned above. 
This can be achieved by redefining
the coefficients in \eqs\nr{Gamma_p} and \nr{Gamma_m} 
and the density matrix as 
\be
 \hat{\Gamma}^{\pm}_{(a\tau)\I\J}(t) \; \equiv \;
  e^{i(\omega^k_\J - \omega^k_\I)t} \, 
 {\Gamma}^{\pm}_{(a\tau)\I\J}
 \;, \quad
 \hat{\rho}^{ }_{\tau\I;\sigma\J}(t) \; \equiv \;
  e^{i(\omega^k_\J - \omega^k_\I)t} \, 
 {\rho}^{ }_{\tau\I;\sigma\J}(t)
 \;. \la{time-dep}
\ee
The evolution equation for ${\rho}$ then 
obtains an additional term of $\rmO(h^0)$, 
\be
 \langle \dot{{\rho}}^{ }_{\tau\I;\sigma\J} \rangle = 
 i (\omega^k_\I - \omega^k_\J)
  \langle {\rho}^{ }_{\tau\I;\sigma\J} \rangle 
 + \rmO(h^2)
 \;, \la{evol_free}
\ee
where the part of $\rmO(h^2)$ has the same structure as in 
\eq\nr{evol_rho} but with time-independent coefficients
(${\Gamma}^\pm$). We remark that in text books density matrices 
are usually defined in terms of ``states''
rather than ``operators'', and then the free evolution has 
the sign of the Liouville - von Neumann equation,   
$i \partial_t {\rho} = [{H}^{ }_0,{\rho}] + ... \;$.
In the present paper we defined the sterile neutrino ``density matrix'' 
through operators, cf.\ \eq\nr{def_rho}. Correspondingly
the free time evolution in \eq\nr{evol_free} has the same  
sign as appears in operator equations of motion, cf.\ \eq\nr{eom_a}.
If desired the sign difference could be eliminated by reversing the ordering 
of indices~\cite{sigl}, however in practice it is inconsequential so we
do not bother.  

%
\subsection{Time evolution of lepton asymmetries}
\la{ss:na}

The derivation of \se\ref{ss:rho}
can be repeated for lepton asymmetries. The starting
point is the equation of motion in \eq\nr{eom_L}, and we take
$n^{ }_a(t) \equiv L^{ }_a(t)/V$ to play the role of $O(t)$
in \eq\nr{eom_general}.
Apart from neutrino Yukawa interactions, 
at high temperatures 
lepton asymmetries are also violated by sphaleron processes. The 
observables not affected by the latter are the linear 
combinations $n^{ }_a - \nB^{ }/{3}$. For these
the final result can be expressed
in close analogy with \eq\nr{evol_rho}, 
\ba
 \Bigl\langle \dot{n}^{ }_a - \frac{\dot{n}^{ }_\rmii{B}}{3} \Bigr\rangle
 & = & 
 \fr12 \int_\vec{k} \sum_{\I,\J,\tau}
 \biggl\{
   \Bigl[
          \hat{\Gamma}^{+}_{(a\tau)\J\I}(t) \, + 
          \hat{\Gamma}^{+*}_{(a\tau)\I\J}(t) \, 
   \Bigr]
   \Bigl[
          \langle \hat{\rho}^{ }_{\tau\I;\tau\J} \rangle
           \; - \;
          \delta^{ }_{\I\J} 
          \, \nF(\omega^k_\J - \mu^{ }_a) 
   \Bigr] 
 \nn 
 & - & 
   \Bigl[
          \hat{\Gamma}^{-}_{(a\tau)\J\I}(t) \, + 
          \hat{\Gamma}^{-*}_{(a\tau)\I\J}(t) \, 
   \Bigr]
   \Bigl[
          \langle \hat{\rho}^{ }_{\tau\I;\tau\J} \rangle
           \; - \;
          \delta^{ }_{\I\J} 
          \, \nF(\omega^k_\J + \mu^{ }_a) 
   \Bigr] 
 \biggr\} + \rmO(h^3)
 \;. \hspace*{9mm} \la{evol_na} 
\ea
The coefficients $\hat{\Gamma}^{\pm}$ are identical to those
in \eq\nr{evol_rho}. There are again two terms, reflecting the 
fact that lepton asymmetry can increase through the production
of leptons or the disappearance of antileptons. 
Eq.~\nr{evol_na}
is odd in charge conjugation, and only the helicity-diagonal
components of the sterile neutrino density matrix contribute.  

%
\subsection{Simplified form of evolution equations}
\la{ss:summary}

We noted in the context of \eq\nr{evol_rho} that non-diagonal helicity 
components of the density matrix decouple from the diagonal ones, 
and in \eq\nr{evol_na} that only the diagonal ones contribute to 
lepton asymmetries.  
Therefore, we omit the non-diagonal components 
in the following. Moreover, for easier inclusion 
of thermal mass corrections, we implement the 
redefinition in \eq\nr{time-dep}, and denote 
\be
 \rho^{ }_{(\tau)\I\J} \equiv \langle {\rho}^{ }_{\tau\I;\tau\J} \rangle
 \;. 
 \la{def_rho_new}
\ee 

The evolution equations in \eqs\nr{evol_rho} and \nr{evol_na} can be  
simplified if we expand the right-hand sides to first order in the lepton
chemical potentials, assuming $\mu^{ }_a, \muB^{ } \ll \pi T$. 
Within perturbation theory the presence of $\mu^{ }_a, \muB^{ } \neq 0$
implies that the temporal component of the hypercharge gauge potential
develops an expectation value, guaranteeing the hypercharge neutrality
of the plasma; this expectation value is conventionally referred to as
the hypercharge chemical potential, denoted by
$\muY$ (cf.\ \se\ref{ss:general_mu}).  
In this limit the coefficients in \eqs\nr{Gamma_p} and \nr{Gamma_m},  
redefined through \eq\nr{time-dep}, have the forms
(cf.\ \se\ref{se:coeffs}; $\bar{\mu}^{ } \equiv \mu^{ } / T$)
\ba
 {\Gamma}^+_{(a\tau)\I\J} & = & 
 h^*_{\I a}h^{ }_{\J a} \, 
 \bigl[
   Q^{ }_{(\tau)\I\J} + \bar{\mu}^{ }_a R^{ }_{(\tau)\I\J}
  + \bmuY^{ } S^{ }_{(\tau)\I\J}
 \bigr] + \rmO(\bar{\mu}^2)
 \;, \la{QRS1} \\ 
 {\Gamma}^-_{(a\tau)\I\J} & = & 
 h^{ }_{\I a}h^{*}_{\J a} \, 
 \bigl[
   Q^{ }_{(-\tau)\I\J} - \bar{\mu}^{ }_a R^{ }_{(-\tau)\I\J}
  - \bmuY^{ } S^{ }_{(-\tau)\I\J}
 \bigr] + \rmO(\bar{\mu}^2)
 \;. \la{QRS2}
\ea
In principle there are also coefficients proportional to 
$ \bmuB^{ } $, appearing like those proportional to 
$ \bmuY^{ } $, however these vanish at leading order 
(cf.\ \se\ref{se:coeffs}) and are omitted here already. 

The functions $Q,R$ and $S$, estimated in \se\ref{se:coeffs}, are 
found to be real. To a reasonable approximation they are also 
symmetric in $I \leftrightarrow J$, however this symmetry is 
broken by the ``soft'' 
$1+n \leftrightarrow 2+n$ scatterings evaluated in 
\se\ref{ss:1to2}. Roughly speaking, the 
coefficient $ {\Gamma}^+_{(a\tau)\I\J} $ describes the amplitude
${}^{ }_T\langle J | I \rangle^{ }_0$, where $|...\rangle^{ }_T$
implies that the state evolves within a medium. 
Even though  
${}^{ }_0\langle J | I \rangle^{ }_0
 = {}^{ }_0\langle I | J \rangle^{*}_0
$, 
it is possible that 
${}^{ }_T\langle J | I \rangle^{ }_0
 \neq {}^{ }_T\langle I | J \rangle^{*}_0
$.

The physical meaning of the equations can be made more transparent by taking
the helicity-symmetric and antisymmetric parts of $\rho^{ }_{(\tau)}$
as the basic variables. Correspondingly we define
\be
 \rho^{\pm}_{\I\J} \; \equiv \; 
 \frac{
   \rho^{ }_{(+)\I\J} \pm \rho^{ }_{(-)\I\J} 
 }{2}
 \;. \la{rho_pm}
\ee
Furthermore, in order to streamline the equations, we make use of the 
kinematic simplification in \eq\nr{ur}. 
This implies that momenta $k \lsim M^{ }_\I \ll \pi T$
are not treated properly, however their contribution to 
lepton asymmetries is strongly phase-space suppressed
($M^{ }_\I \sim 10^{-2} \pi T$). 

Let us first inspect the equation for lepton asymmetries, 
\eq\nr{evol_na}. Inserting \eqs\nr{ur}, \nr{QRS1} and \nr{QRS2}
into \eq\nr{evol_na} we obtain
\be
 \Bigl\langle \dot{n}^{ }_a - \frac{\dot{n}^{ }_\rmii{B}}{3} 
 \Bigr\rangle \; = \;  
 4 \int_\vec{k} 
 \tr \Bigl\{ 
  - 
  \, \nF(k) [1 - \nF(k)] \,  A^{+}_{(a)}
   +   
   \bigl[\, \rho^{+}_{ } - \unit \, \nF(k) \, \bigr]
  B^{+}_{(a)} + \rho^{-}_{ } B^{-}_{(a)}
 \Bigr\} 
 \; + \; \rmO(\mu_a^2) 
 \;, 
 \la{summary_na}  
\ee
where 
\ba
 A^{+}_{(a)\I\J}
 & \equiv & 
  \re (h^{ }_{\I a}h^*_{\J a}) \, 
 \bar{\mu}^{ }_a \,
 Q^{+}_{\{\I\J\}} 
 \;, \la{A} \\[2mm] 
 B^{+}_{(a)\I\J}
 & \equiv & 
 - i \im (h^{ }_{\I a}h^*_{\J a}) \, 
    Q^{+}_{\{\I\J\}} 
  + 
  \re (h^{ }_{\I a}h^*_{\J a}) \, 
  \Bigl[ 
  \bar{\mu}^{ }_a \, 
        R^{+}_{\{ \I \J\} } 
   + \bmuY^{ } \,
        S^{+}_{\{ \I\J\} } 
  \Bigr]
 \;, \la{Bplus} \\
 B^{-}_{(a)\I\J}
  & \equiv &
  \re (h^{ }_{\I a}h^*_{\J a}) \, 
    Q^{-}_{\{\I\J\}} 
 - i \im (h^{ }_{\I a}h^*_{\J a}) \, 
  \Bigl[ 
    \bar{\mu}^{ }_a \, 
     R^{-}_{\{ \I \J\} } 
   + \bmuY^{ } \, 
        S^{-}_{\{ \I\J\} }
  \Bigr]
 \;. \la{Bminus}
\ea
Here 
\be
 Q^{\pm}_{\I\J} \; \equiv \; 
 \frac{ Q^{ }_{(+)\I\J} \pm
 Q^{ }_{(-)\I\J} }{2}  
 \;, \quad
 Q^{\pm}_{\{\I\J\}} \; \equiv \; 
 \frac{ Q^{\pm }_{\I\J} + Q^{\pm}_{\J\I} }{2}
 \la{symm}
\ee
denote 
a symmetrization or antisymmetrization
over the helicity-flipping and conserving 
contributions (cf.\ \se\ref{se:coeffs}), and 
a symmetrization over the generation indices, respectively.  

The first term 
on the right-hand side of \eq\nr{summary_na}
is a ``washout term'', decreasing any lepton
asymmetry towards zero. It agrees with the corresponding term 
derived from different considerations in ref.~\cite{washout}.  
The second and third terms 
are ``source terms'', generating a lepton asymmetry.
They display a product of structures manifesting Sakharov-type conditions, 
namely 
deviation from thermal equilibrium
and CP violation.
One of the sources originates from a helicity-symmetric  
$\rho^{ }_{ }$ and the other from 
a helicity-asymmetric one.
The parts proportional to chemical potentials in \eqs\nr{Bplus} and \nr{Bminus}
are of second order in the language of linear response theory, 
containing a product of two deviations from equilibrium 
(chemical potentials and a non-thermal
density matrix). We include them in the equations, given that 
the density matrix may deviate significantly from equilibrium.

Analogous equations are obtained for the density matrices. Displaying
them as a pair of complex Hermitean matrices with generation 
indices, we obtain
\ba
  \dot{\rho}^{+}_{ }  & = & 
   i \bigl[H^{ }_{0}, \rho^{+}_{ } \bigr] 
   + 
   i \bigl[\Delta^{ }_{0}, \rho^{-}_{ } \bigr] 
   + 2 \nF(k) [ 1-  \nF(k)] \, C^{+}_{ } \nn[2mm] 
   & - & 
   D^{+}_{ }  
   \bigl[ \rho^{+}_{ } - \unit \nF(k) \bigr]
  - 
   \bigl[ \rho^{+}_{ } - \unit \nF(k) \bigr]  
   D^{+\dagger}_{ }
  - D^{-}_{ }  \rho^{-}_{ }
  - \rho^{-}_{ }  D^{-\dagger}_{ }
  \; + \; \rmO(\mu_a^2)
 \;, \la{summary_rho_plus} \\[2mm]
  \dot{\rho}^{-}_{ }  & = & 
   i \bigl[H^{ }_{0}, \rho^{-}_{ } \bigr] 
  +  i \bigl[\Delta^{ }_{0}, \rho^{+}_{ } \bigr] 
   + 2 \nF(k) [ 1-  \nF(k)] \, C^{-}_{ } \nn[2mm] 
   & - & 
   D^{-}_{ }  
   \bigl[ \rho^{+}_{ } - \unit \nF(k) \bigr]
  - 
   \bigl[ \rho^{+}_{ } - \unit \nF(k) \bigr]  
   D^{-\dagger}_{ }
  - D^{+}_{ }  \rho^{-}_{ }
  - \rho^{-}_{ }  D^{+\dagger}_{ }
 \; + \; \rmO(\mu_a^2)
 \;. \la{summary_rho_minus}
\ea
The coefficient matrices read
\ba
 C^{+}_{\I\J} & \equiv & 
 - i {\textstyle\sum_a} 
   \im (h^{ }_{\I a} h^{*}_{\J a})\, \bar{\mu}^{ }_a  
   \, Q^{+}_{\{\I\J\} }
 \;, \la{Cplus} \\[2mm] 
 C^{-}_{\I\J} & \equiv & 
 {\textstyle\sum_a} 
 \re (h^{ }_{\I a} h^{*}_{\J a})\, \bar{\mu}^{ }_a  
   \, Q^{-}_{\{\I\J\} }
 \;, \la{Cminus} \\[2mm] 
 D^{+}_{\I\J} & \equiv & 
 {\textstyle\sum_a}
   \re (h^{ }_{\I a} h^{*}_{\J a})   
   \,  
   Q^{+}_{\I\J}
  -  i \, {\textstyle\sum_a} \im (h^{ }_{\I a} h^{*}_{\J a})  \, 
  \Bigl[ \bar{\mu}^{ }_a \, R^{+}_{\I\J}
   + \bmuY^{ } S^{+}_{\I\J}
  \Bigr] \,  
 \;, \la{Dplus}  \\[2mm]
 D^{-}_{\I\J} & \equiv & 
  -  i \, {\textstyle\sum_a} \im (h^{ }_{\I a} h^{*}_{\J a})  \, 
   Q^{-}_{\I\J}
  + 
 {\textstyle\sum_a}
   \re (h^{ }_{\I a} h^{*}_{\J a})   
   \,  
  \Bigl[ \bar{\mu}^{ }_a \, R^{-}_{\I\J}
   + \bmuY^{ } S^{-}_{\I\J}
  \Bigr]
 \;. \la{Dminus}  
\ea
The coefficients  $ C^{\pm}_{ } $ 
generate non-thermal density matrices if lepton asymmetries
are present. 
The coefficients $D^{\pm}_{ }$ are ``washout terms'' in the sense 
that they drive the system towards equilibrium, but they are often called
``production rates'', because normally 
$\rho^{+}_{ } \ll \mathbbm{1} \nF{}(k)$. 
Then $D^{+}_{ }$ produces $\rho^{+}_{ }$ 
and $D^{-}_{ }$ produces $\rho^{-}_{ }$.   
In the limit of 
a single generation 
$D^{+}_{ }$ agrees with a term derived from 
different considerations in refs.~\cite{dmpheno,equil}.

The Hermitean matrices $H^{ }_0$ and $\Delta^{ }_0$
on the first lines of \eqs\nr{summary_rho_plus} and \nr{summary_rho_minus} 
contain the vacuum energies but also the thermal 
mass corrections (cf.\ appendix~\ref{ss:mass}), 
\ba
 H^{ }_{0\I\J} & = & k\, \delta^{ }_{\I\J} + \frac{1}{2k} \biggl[
 \delta^{ }_{\I\J} M_\I^2
 + 
 \frac{{\textstyle\sum_a} \re (
     h^{ }_{\I a}h^{*}_{\J a} ) T^2
   }{4}  \biggr]  
 + \rmO\Bigl(\frac{1}{k^3}\Bigr)
 \;, \la{H0} \\[2mm]
 \Delta^{ }_{0\I\J} & = &  
 - 
 \frac{i\, {\textstyle\sum_a} \im(
     h^{ }_{\I a}h^{*}_{\J a}  ) T^2
   }{8k}    
 + \rmO\Bigl(\frac{1}{k^3}\Bigr)
 \;. \la{Delta0}
\ea
The first term in \eq\nr{H0}, proportional to the unit matrix, drops
out in the commutators in \eqs\nr{summary_rho_plus} and 
\nr{summary_rho_minus}.
Solving the time evolution with $H^{ }_0$ and $\Delta^{ }_0$ 
implements a resummation of thermal mass corrections,\footnote{%
 The thermal mass squared appearing here is the so-called
 ``asymptotic'' mass, relevant for $k \gsim \pi T$; 
 it is twice as large as the ``soft'' thermal mass squared, 
 relevant for $k \ll \pi T$~\cite{weldon}. 
 } 
avoiding dispersive secular terms in the evolution equations. 
It may be wondered whether the equilibrium terms appearing in 
\eqs\nr{summary_na}, \nr{summary_rho_plus} and \nr{summary_rho_minus}, 
involving the Fermi 
distribution, should also contain the eigenvalues of the 
system described by $H^{ }_0$ and $\Delta^{ }_0$ as 
arguments. This is, however,  
a higher-order effect in the ultrarelativistic
regime $k \sim \pi T \gg M^{ }_{\I}$.

We conclude by remarking that a set of equations similar
to \eqs\nr{summary_na}, \nr{summary_rho_plus} and \nr{summary_rho_minus}
was obtained in ref.~\cite{n3}, however without the inclusion of the
hypercharge chemical potential~$\bmuY^{ }$ (and hence of the IR sensitivities 
discussed in \se\ref{se:coeffs}) and under the assumption 
that terms of order $\bar{\mu}^{ }_a\, \rho^{-}_{ }$
could be dropped, as they represent a double deviation from equilibrium.
Moreover helicity conserving contributions  
were neglected together with the generation ($IJ$) dependence 
of the helicity flipping ones. 
As discussed in \se\ref{se:coeffs} and confirmed numerically 
in \se\ref{ss:numerics},
the latter assumptions hold well for $M^{ }_{\I}\ll T$. 
Specifically, upon setting 
$\bmuY^{ }\to 0$, $Q^{ }_{(-)\I\J}\to 0$, $ R^{ }_{(-)\I\J}\to 0$, 
$Q^{ }_{(+)\I\J}\to \gamma^{(0)}$,
$R^{ }_{(+)\I\J}\to \gamma^{(2)}$, 
dropping terms proportional to $\bar{\mu}^{ }_a\,\rho^-$, 
and recalling the discussion below \eq\nr{evol_free} concerning the sign
of the commutator terms, 
we can reproduce \eqs(2.16) and (2.19) of ref.~\cite{n3}.

%
\subsection{Rate for fermion number non-conservation}
\la{ss:fermion}

It is well-known that if all Majorana masses are set to zero 
in \eq\nr{L_Majorana}, then the theory has an additional conserved charged, 
which we may call the ``fermion number''. Indeed the Majorana spinor
can be replaced by a chiral Dirac spinor in this case, and the 
conserved charge then counts the total asymmetry in 
right-handed and left-handed leptons. Keeping instead the Majorana
character intact, the fermion number is defined as the sum of the helicity 
asymmetry of the sterile neutrinos and the lepton asymmetry of 
the Standard Model sector. It is easy to 
demonstrate that \eqs\nr{summary_na} and 
\nr{summary_rho_minus} respect this symmetry for $M^{ }_{\I}/T\to 0$. 
A straightforward computation yields
\ba
 && \hspace*{-1.5cm} 
 \sum_a \Bigl\langle \dot{n}^{ }_a - \frac{\dot{n}^{ }_\rmii{B}}{3} 
 \Bigr\rangle + 2 \int_\vec{k } \tr \bigl( \dot{\rho}^{-}_{ } \bigr)
 \nn[2mm]
 & = & 
 4 \int_\vec{k} \sum_a \tr 
 \Bigl\{ 
   - \nF(k) [ 1-  \nF(k)] E^{ }_{(a)}
   + \bigl[ \rho^{+}_{ } - \unit \nF(k) \bigr] F^{ }_{(a)}
   - \rho^{-}_{ } G^{ }_{(a)} 
 \Bigr\} 
 \;, \la{fermion} 
\ea
where the coefficients read
\ba
 E^{ }_{(a)\I\J} & \equiv & 
 \re (h^{ }_{\I a} h^{*}_{\J a})\, \bar{\mu}^{ }_a  
   \, Q^{ }_{(-)\{\I\J\} }
 \;, \la{E} \\[2mm]
 F^{ }_{(a)\I\J} & \equiv & 
  -  i \im (h^{ }_{\I a} h^{*}_{\J a})  \, 
   Q^{ }_{(-)\{\I\J\}}
  + 
   \re (h^{ }_{\I a} h^{*}_{\J a})   
   \,  
  \Bigl[ \bar{\mu}^{ }_a \, R^{ }_{(-)\{\I\J\}}
   + \bmuY^{ } S^{ }_{(-)\{\I\J\}}
  \Bigr]
 \;, \la{F} \\[2mm]
 G^{ }_{(a)\I\J} & \equiv & 
   \re (h^{ }_{\I a} h^{*}_{\J a})   
   \,  
   Q^{ }_{(-)\{\I\J\}}
  -  i \im (h^{ }_{\I a} h^{*}_{\J a})  \, 
  \Bigl[ \bar{\mu}^{ }_a \, R^{ }_{(-)\{\I\J\}}
   + \bmuY^{ } S^{ }_{(-)\{\I\J\}}
  \Bigr] 
 \;. \la{G}
\ea
All contributions are proportional to 
the helicity-conserving coefficients 
$ Q^{ }_{(-)\{\I\J\}} $, 
$ R^{ }_{(-)\{\I\J\}} $, or 
$ S^{ }_{(-)\{\I\J\}} $. 
As demonstrated in \se\ref{se:coeffs} 
these are suppressed
by $\sim M^{ }_{\I}/(gT)$ in comparison with 
helicity-flipping coefficients, and are in general numerically 
insignificant (cf.\ \figs\ref{fig:coeffQ}--\ref{fig:coeffS}), 
save for the fact that due to their infrared sensitivity
they peak around the electroweak crossover. 

%
\section{Determination of coefficient functions $Q,R$ and $S$}
\la{se:coeffs}

%
\subsection{General setup}

In order to determine the functions $Q,R$ and $S$ defined
in \eqs\nr{QRS1} and \nr{QRS2} which parametrize 
\eqs\nr{A}--\nr{Bminus}, \nr{Cplus}--\nr{Dminus}, \nr{E}--\nr{G},  
we need to evaluate the amplitudes
\be
 \bar{u}^{ }_{\vec{k}\tau\J} 
 \, \aL \, \rho^{ }_a(\mathcal{K}^{ }_\J) \, \aR \,   
 u^{ }_{\vec{k}\tau\I}
 \;, \quad
 \bar{u}^{ }_{\vec{k}\tau\J} 
 \, \aR \, \rho^{ }_a( - \mathcal{K}^{ }_\J) \, \aL \,   
 u^{ }_{\vec{k}\tau\I}
 \;, \la{M}
\ee 
cf.\ \eqs\nr{Gamma_p} and \nr{Gamma_m}.
Here the spectral function $\rho^{ }_a$ corresponds to 
the operators in \eq\nr{ja}. 
To zeroth order in chemical potentials, a fermionic spectral function
is even in its argument: 
$
 \rho^{ }_a( - \mathcal{K}^{ }_\J) = 
 \rho^{ }_a( \mathcal{K}^{ }_\J) + \rmO(\mu)
$.
This explains a part of the properties in 
\eqs\nr{QRS1} and \nr{QRS2}, but the dependence on 
the helicity $\tau$ remains to be worked out. 

For determining the dependence on $\tau$,
the form of the spinor $u^{ }_{\vec{k}\tau\I}$ is needed. 
We can write
\be
 u^{ }_{\vec{k}\tau\I} \; = \; 
 \frac{\bsl{\mathcal{K}}^{ }_{\!\I} + M^{ }_\I}{\sqrt{\omega^k_\I + M^{ }_\I}}
 \, \eta^{ }_\tau
 \;, \la{u} 
\ee
where the spinors satisfy
$
 \sum_{\tau = \pm} \eta^{ }_\tau \bar{\eta}^{ }_{\tau} = \fr12 
 (\unit + \gamma^0)
$. 
The precise form of $\eta^{ }_\tau$ depends on the representation
chosen for the Dirac matrices. Examples for 
the standard and Weyl representations are
\be
 \gamma^0 
 = 
 \left( 
   \begin{array}{cc} 
     \unit^{ }_{ } & 0 \\ 
     0 & \! -\unit^{ }_{ }
   \end{array}
 \right)
 \; \Rightarrow \; 
 \eta^{ }_\tau = 
 \left( \begin{array}{c} 
   | \tau \rangle \\ 
   0 
  \end{array}
 \right)
 \;, \quad
 \gamma^0 
 = 
 \left( 
   \begin{array}{cc} 
     0 & \unit^{ }_{ } \\ 
     \unit^{ }_{ } & 0
   \end{array}
 \right)
 \; \Rightarrow \; 
 \eta^{ }_\tau = \frac{1}{\sqrt{2}}
 \left( \begin{array}{c} 
   | \tau \rangle \\ 
   | \tau \rangle 
  \end{array}
 \right)
 \;. 
\ee
Here $|\tau\rangle$ is a helicity eigenstate, 
$\sum_i k^i {\sigma}^{ }_i |\tau \rangle  = 
\tau |\vec{k}| |\tau \rangle$, where
$\sigma^{ }_i$ are the Pauli matrices.

%
\subsection{$1+n \leftrightarrow 2+n$ scatterings}
\la{ss:1to2}

%
\begin{figure}[t]
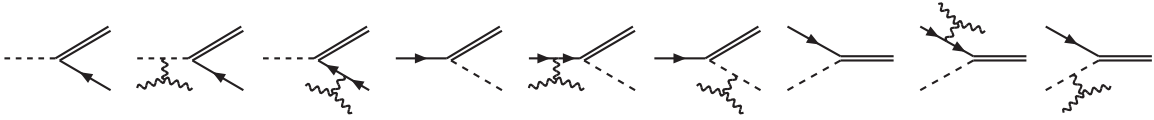


\begin{eqnarray*}
&& 
 \hspace*{-1.0cm}
 \xproc
 \hspace*{0.50cm}
 \xprocG
 \hspace*{0.40cm}
 \xprocH
 \hspace*{0.50cm}
 \yproc
 \hspace*{0.50cm}
 \yprocG
 \hspace*{0.40cm}
 \yprocH
 \hspace*{0.50cm}
 \zproc
 \hspace*{0.50cm}
 \zprocG
 \hspace*{0.40cm}
 \zprocH
\end{eqnarray*}


\caption[a]{\small 
 Examples of $1+n\leftrightarrow 2+n$ scatterings 
 contributing to the sterile neutrino spectral function
 (the spectral function is a cut, i.e.\ ``amplitude squared''
 of such processes, 
 convoluted with the appropriate distribution functions). 
 Sterile neutrinos are denoted by a double line, whereas 
 arrowed, dashed, and wiggly lines correspond to Standard Model 
 fermions, scalars, and gauge fields, respectively.
} 
\la{fig:1to2}
\end{figure}
%

Physically, a non-vanishing spectral weight originates
as a result of scatterings.  The operator $j^{ }_a$ in \eq\nr{ja}
couples directly
to two fields, the Higgs doublet and active lepton doublets, and if
no further particles are involved we call the process a 
$1 \leftrightarrow 2$ scattering
(cf.\ \fig\ref{fig:1to2}). 
Such scatterings give no contribution
in the massless limit, because there is no phase space for the 
on-shell process. In the presence of thermal masses and Majorana
masses, one of the kinematic channels may open up. If we count the masses
as being of order $M^{ }_\I\sim m^{ }_\phi \sim m^{ }_{\ell} \sim gT$, 
then this contribution, 
suppressed by the small masses, is parametrically of the same
order as that of $2\leftrightarrow 2$ scatterings
(cf.\ \fig\ref{fig:2to2}). 

Given that the masses are small and that thermal momenta
are of order $k\sim \pi T$, the computation of $1\leftrightarrow 2$
scatterings can be simplified by considering  
ultrarelativistic kinematics, cf.\ \eq\nr{ur}. 
However, there is also a complication, namely that soft scatterings
which do not modify the kinematics are not 
suppressed and need to be resummed to all orders. This so-called 
Landau-Pomeranchuk-Migdal (LPM) resummation was worked 
out in ref.~\cite{bb1}. We need to generalize these results, 
because in ref.~\cite{bb1} a sum was taken over the two helicity
states, and because only the diagonal elements
of the density matrix were considered, and because 
the lepton chemical potentials were set to zero.\footnote{%
 In ref.~\cite{n3} $\mu^{ }_a$ was included but the 
 hypercharge chemical potential was omitted.}

It is not too difficult to generalize the results of ref.~\cite{bb1} 
to apply to the spectral function. 
Adopting the notation in \se{3.1} of ref.~\cite{broken} and noting
that all thermal masses 
are even in $\mu^{ }_a$, because they represent elastic scatterings 
through gauge exchange which can be both off fermions and off antifermions, 
the resummed ``helicity-conserving'' and 
``helicity-flipping'' wave functions
are denoted by $g^{ }_\J$ and $\vec{f}^{ }_\J$. The latter is a $p$-wave
(vector) object, and it flips the helicity from that carried by Standard
Model leptons. The spectral function can be expressed in terms of these as 
\ba
 \rho^{ }_a(\mathcal{K}^{ }_\J)^\rmii{LPM}_{ }
 & \equiv & 
  \frac{1}{4\pi} 
 \int_{-\infty}^{\infty} \! {\rm d}\omega_1 \, 
 \int_{-\infty}^{\infty} \! {\rm d}\omega_2 \;
 \delta(\omega^k_\J - \omega_1 - \omega_2) \, 
 \bigl[ 1 - \nF(\omega_1 - \muLa)
  + \nB(\omega_2 - \muH^{ }) \bigr] 
 \nn  
 & \times & 
  \, \frac{1}{\omega_2} 
 \lim_{\vec{y} \to \vec{0}}
 \biggl\{
   \biggl( \gamma^0 - \frac{\vec{k}\cdot \bm{\gamma}}{k} \biggr)
   \im\, \bigl[g^{ }_\J(\vec{y} )\bigr]  + 
   \frac{\gamma^0}{2 \omega_1^2}
   \im\, \bigl[\nabla_\perp\cdot \vec{f}^{ }_\J(\vec{y} )\bigr] 
 \biggr\}
 \; + \; \rmO\Bigl( \frac{g^3 T}{\pi^3} \Bigr) 
 \;. \nn \la{lpm_symm}
\ea
Here the chemical potentials are
$\muLa \equiv \mu^{ }_a - \muY^{ }/2$
and $\muH^{ } \equiv \muY^{ }/2$,  
where $\muY^{ }$ is the hypercharge chemical potential
(the expression of $\muY^{ }$ in terms of the $\mu^{ }_a$
is recalled in \se\ref{se:mu}).

Taking subsequently projections such as in \eq\nr{M}
(cf.\ \eq\nr{trace2});
expanding as in \eq\nr{ur}; and employing $k$ 
as a variable instead of $\omega^k_\J$, whereby
terms suppressed by $\rmO(M_\I^2 / k^2)$ are omitted,  
we find
\ba
 & & \hspace*{-1.5cm}
 \bar{u}^{ }_{\vec{k}\tau\J} 
 \,\aL \,  \rho^\rmii{LPM}_a(\pm \mathcal{K}^{ }_\J) \, \aR \, 
 u^{ }_{\vec{k}\tau\I} 
 \nn[2mm] 
 & = & 
  \frac{1}{4\pi} 
 \int_{-\infty}^{\infty} \! {\rm d}\omega_1 \, 
 \int_{-\infty}^{\infty} \! {\rm d}\omega_2 \;
 \delta(k - \omega_1 - \omega_2) \, 
 \bigl[ 1 - \nF(\omega_1 \mp \muLa)
  + \nB(\omega_2 \mp \muH^{ }) \bigr] 
 \nn  
 & \times & 
  \, \frac{k}{\omega_2} 
 \lim_{\vec{y} \to \vec{0}}
 \biggl\{
    \frac{M^{ }_\I M^{ }_\J\, \delta^{ }_{\tau,-} }{k^2}
   \im\, \bigl[g^{ }_\J(\vec{y} )\bigr]
   \;  +  \; \frac{ \delta^{ }_{\tau,+} }{\omega_1^2}
   \im\, \bigl[\nabla_\perp\cdot \vec{f}^{ }_\J(\vec{y} )\bigr] 
 \biggr\}
 \; + \; \rmO\Bigl( \frac{g^3 T^2}{\pi^2} \Bigr)
 \;. \la{rho_LPM}
\ea
For the latter chiral projection in \eq\nr{M}, 
$ \delta^{ }_{\tau,-} $ and $ \delta^{ }_{\tau,+} $ 
are exchanged.
The coefficients $Q,R,S$ in \eqs\nr{QRS1} and \nr{QRS2} are
obtained by making the following substitutions in \eq\nr{rho_LPM} 
(expanding again $1/\sqrt{\omega^k_\I \omega^k_\J} \approx 1/k$
in \eqs\nr{Gamma_p} and \nr{Gamma_m}): 
\ba
 \bigl[ 1 - \nF(\omega_1 \mp \muLa)
  + \nB(\omega_2 \mp \muH^{ }) \bigr] 
 & \stackrel{Q^\rmii{LPM}}{\longrightarrow} & 
 \frac{
 1 - \nF(\omega_1 )
  + \nB(\omega_2 )
 }{k}
 \;, \la{Q12} \\ 
 & \stackrel{R^\rmii{LPM}}{\longrightarrow} & 
 \frac{
   T \nF'(\omega_1 ) 
 }{k}
 \;, \hspace*{5mm} \\ 
 & \stackrel{S^\rmii{LPM}}{\longrightarrow} & 
  - \frac{
   T \nF'(\omega_1 ) 
   + T \nB'(\omega_2 ) 
 }{2 k}
  \;. \hspace*{5mm} \la{S12}
\ea 
Here 
$
 T \nF'(\omega_1) = - \nF(\omega_1) [1 - \nF(\omega_1)] 
$ 
 and
$
 T \nB'(\omega_2) = - \nB(\omega_2) [1 + \nB(\omega_2)]  
$.\footnote{%
 The weight $\nB'(\omega_2)$ is quadratically divergent around $\omega_2 = 0$.
 In the context of \eq\nr{rho_LPM} this leads to a linear divergence, however
 the integral is well-defined as a principal value. 
 }
The Hamiltonian is also written in terms of $k$, 
\be
 \hat{H}^{ }_\J \; \equiv \; - \frac{M^2_\J}{2 k} + 
 \frac{m_{\ell}^2 - \nabla_\perp^2}{2\omega_1} + 
 \frac{m_\phi^2 - \nabla_\perp^2}{2\omega_2} - i \, \Gamma(y) 
 \quad y \equiv |\vec{y}^{ }_\perp| 
 \;, \la{HJ}
\ee
where $\Gamma(y)$ is given in \eq(3.3) 
of ref.~\cite{broken}, 
$
 m^2_{\ell} \equiv \lim_{\mu^{ }_a \to 0} m^2_{\ell,a} 
$ is given in 
\eq\nr{mell}, and 
\be
  m_\phi^2 =  -\frac{\mH^2}{2} + 
  \Bigl( g_1^2 + 3 g_2^2 + 4 h_t^2  + 8 \lambda  
  \Bigr) \frac{T^2}{16}
 \;, \la{mphi}
\ee
where $\mH$ is the physical Higgs mass. 
The wave functions are obtained from
\be
 (\hat{H}^{ }_\J + i 0^+)\, g^{ }_\J(\vec{y}) \, = \, 
  \delta^{(2)}(\vec{y}) \;, \quad 
 (\hat{H}^{ }_\J + i 0^+)\, \vec{f}^{ }_\J(\vec{y}) \, = \, 
  -\nabla^{ }_\perp \delta^{(2)}(\vec{y}) 
 \;. \la{Seq}
\ee
For the numerical solution we adopt the procedure
described in ref.~\cite{interpolation}; at $T \lsim 160$~GeV, 
when we are in the Higgs phase, 
it needs to be modified as explained in ref.~\cite{broken}. 

%
\subsection{Hard $2 \leftrightarrow 2$ scatterings}
\la{ss:22}

%
\begin{figure}[t]
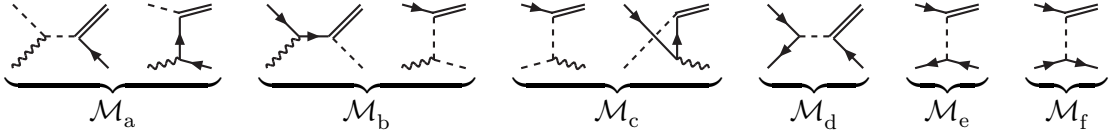


\begin{eqnarray*}
&& 
 \hspace*{-1cm}
 \underbrace{
 \procGa
 \hspace*{-0.00cm}
 \procHa
 \hspace*{-0.20cm}
  }_{\textstyle\mathcal{M}^{ }_\rmi{a}}
 \hspace*{0.45cm}
 \underbrace{
 \procG
 \hspace*{-0.00cm}
 \procH
 \hspace*{-0.20cm}
  }_{\textstyle\mathcal{M}^{ }_\rmi{b}}
 \hspace*{0.45cm}
 \underbrace{
 \procJ
 \hspace*{0.10cm}
 \procI
 \hspace*{-0.40cm}
  }_{\textstyle\mathcal{M}^{ }_\rmi{c}}
 \hspace*{0.45cm}
 \underbrace{
 \procKa
 \hspace*{-0.35cm}
  }_{\textstyle\mathcal{M}^{ }_\rmi{d}}
 \hspace*{0.45cm}
 \underbrace{
 \procKc
 \hspace*{-0.20cm}
  }_{\textstyle\mathcal{M}^{ }_\rmi{e}}
 \hspace*{0.45cm}
 \underbrace{
 \procKb
 \hspace*{-0.20cm}
  }_{\textstyle\mathcal{M}^{ }_\rmi{f}}
\end{eqnarray*}

\vspace*{-4mm}

\caption[a]{\small 
 $2\leftrightarrow 2$ scattering contributions to the 
 sterile neutrino spectral function, cf.\ \eq\nr{boltzmann}
 (the spectral function is a cut, i.e.\ ``amplitude squared'', 
 convoluted with the appropriate distribution functions).  
 The notation is as in \fig\ref{fig:1to2}. 
} 
\la{fig:2to2}
\end{figure}
%

Unlike the $1\leftrightarrow 2$ processes, the $2\leftrightarrow 2$
scatterings, illustrated in \fig\ref{fig:2to2},  
are not phase-space suppressed. Therefore they can be 
computed at leading order in an expansion in $M^{2}_\I / k^2$, i.e.\, 
with massless right-handed neutrinos. Then only 
one helicity state contributes,\footnote{%
 This corresponds to a helicity-flipping process: 
 in the case of gauge scatterings, the angular momentum is supplied
 by a vector particle in the initial or final state. 
 In the case of Yukawa scatterings, there is a left and right-handed
 top quark involved, and their angular momenta balance against 
 those in the lepton sector. 
 } 
and we can write (cf.\ \eq\nr{trace2})
\ba
 \bar{u}^{ }_{\vec{k}\tau\J} 
 \, \aL \, \rho^{2\leftrightarrow 2}_a(\mathcal{K}^{ }_\J) \, \aR \, 
 u^{ }_{\vec{k}\tau\I} 
 & = & 
 \delta^{ }_{\tau,+}
 \sum_{\tau}
 \bar{u}^{ }_{\vec{k}\tau\J} 
 \, \aL \,  \rho^{2\leftrightarrow 2}_a(\mathcal{K}^{ }_\J) \,  \aR \,
 u^{ }_{\vec{k}\tau\J}
 \nn  
 & = & 
 \delta^{ }_{\tau,+}
 \tr \{
  \bsl{\mathcal{K}}^{ }_{\!\J}  
  \, \aL \,  \rho^{2\leftrightarrow 2}_a(\mathcal{K}^{ }_\J) \,  \aR \,
 \}
 \;.  
\ea
For the latter chiral structure in \eq\nr{M}, 
$  \delta^{ }_{\tau,+} $ 
gets replaced with 
$  \delta^{ }_{\tau,-} $. 
Furthermore we can 
replace $\mathcal{K}^{ }_\J$ through
$\mathcal{K} \equiv (k,\vec{k})$, whereby the result is 
independent of the indices $I$ and $J$. 

The $2\leftrightarrow 2$ scatterings contain two logarithmic IR
divergences, related to soft lepton exchange and scattering
off soft Higgs particles, respectively. Following ref.~\cite{bb2},
we handle the former by first carrying out a massless computation
(this subsection),
and subsequently correct the soft exchange contribution through an 
appropriate resummation (\se\ref{ss:lepton_exchange}). 
We denote the unresummed contribution from hard 
momentum exchange by $ \rho^{2\leftrightarrow 2,\rmi{hard}}_a(\mathcal{K}) $.
The latter divergence concerns terms proportional to $\muY$ and was
not present in ref.~\cite{bb2}, however an analogous procedure of
``subtraction'' and ``correction'' 
can be adopted (\se\ref{ss:external_Higgs}). 
The phase space regions from which the divergences originate 
are illustrated in \fig\ref{fig:phasespace}.

\begin{figure}[t]

\hspace*{-0.1cm}
\centerline{%
 \epsfxsize=7.5cm\epsfbox{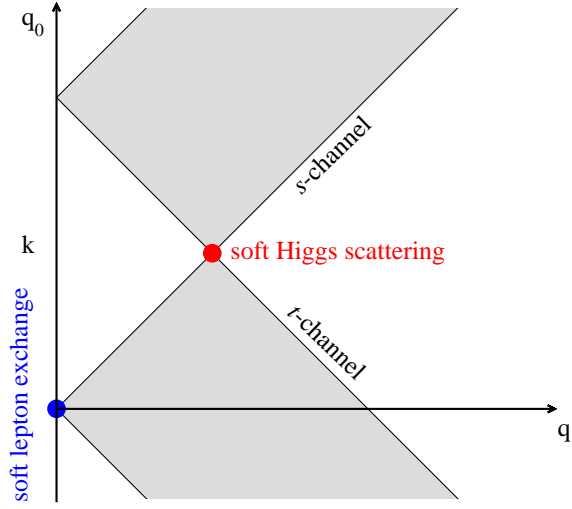}%
}

\caption[a]{\small
 The phase space regions contributing to \eq\nr{2to2}. Here $(q,q^{ }_0)$
 parametrizes the four-momentum of the exchanged particle. 
 There are logarithmic infrared 
 divergences associated with the fermionic  
 contributions $ \Phi_\rmi{$s$f} $ and $ \Phi_\rmi{$t$f} $, 
 from soft lepton exchange (exchanged particle has $(q,q^{ }_0)\approx (0,0)$)
 and soft Higgs scattering (exchanged particle has $(q,q^{ }_0)\approx (k,k)$
 whereas external scatterer is soft). The divergences
 can be resummed as explained in \ses\ref{ss:lepton_exchange} and 
 \ref{ss:external_Higgs}, respectively.  
}

\la{fig:phasespace}
\end{figure}

Accounting for the processes shown in \fig\ref{fig:2to2}, we obtain
\ba
 && \hspace*{-1.5cm}
 \tr \{
  \bsl{\mathcal{K}} 
  \, \aL \,  \rho^{2\leftrightarrow 2,\rmi{hard}}_a(\mathcal{K}) \,  \aR \,
 \} 
 \; = \; \fr12
 \int \! {\rm d}\Omega^{ }_{2\to 2} \; \mathcal{I} \;, 
 \nn 
 \mathcal{I}
 \!\! & = & \!\! 
 \Bigl\{ 
  \nF(k^{ }_1 \! + \! \muLa)
  \bigl[ 1  +  \nB(p^{ }_1) +  \nB(p^{ }_2 \! - \! \muH) 
  \bigr] + \nB(p^{ }_1) \nB(p^{ }_2 \! - \! \muH)
 \Bigl\} 
  \, |\mathcal{M}^{ }_\rmi{a}|^2 
 \nn \!\! & + & \!\!
 \Bigl\{ 
   \nB(k^{ }_1 \! + \! \muH)
   \bigl[ 1 + 
   \nB(p^{ }_1) - \nF(p^{ }_2 \! - \! \muLa) \bigr]
   + \nB(p^{ }_1) \nF(p^{ }_2 \! - \! \muLa)
 \Bigl\} 
  \, |\mathcal{M}^{ }_\rmi{b}|^2 
 \nn \!\! & + & \!\! 
 \Bigl\{ 
   \nB(k^{ }_1) \bigl[1 + 
   \nB(p^{ }_1 \! - \! \muH ) - \nF(p^{ }_2 \! - \! \muLa) \,  \bigr]
   + \nB(p^{ }_1 \! - \! \muH ) \nF(p^{ }_2 \! - \! \muLa)
 \Bigl\} 
  \, |\mathcal{M}^{ }_\rmi{c}|^2 
 \nn \!\! & + & \!\!
 \Bigl\{ 
   \nF(k^{ }_1 \! + \! \muLa) \bigl[1
   - \nF(p^{ }_1 \! + \! \mutL )
   - \nF(p^{ }_2 \! - \! \mutR )
   \bigr] + 
   \nF(p^{ }_1 \! + \! \mutL ) \nF(p^{ }_2 \! - \! \mutR ) 
 \Bigl\} 
  \, |\mathcal{M}^{ }_\rmi{d}|^2 
 \nn \!\! & + & \!\! 
 \Bigl\{ 
   \nF(k^{ }_1 \! + \! \mutR) \bigl[1
   - \nF(p^{ }_1 \! + \! \mutL)
   - \nF(p^{ }_2 \! - \! \muLa)
   \bigr]
  +
   \nF(p^{ }_1 \! + \! \mutL)
   \nF(p^{ }_2 \! - \! \muLa)
 \Bigl\} 
  \, |\mathcal{M}^{ }_\rmi{e}|^2 
 \nn \!\! & + & \!\!  
 \Bigl\{ 
   \nF(k^{ }_1 \! - \! \mutL) \bigl[1
  - \nF(p^{ }_1 \! - \! \mutR)
  - \nF(p^{ }_2 \! - \! \muLa)
   \bigr]
  +
   \nF(p^{ }_1 \! - \! \mutR)
   \nF(p^{ }_2 \! - \! \muLa)
 \Bigl\} 
  \, |\mathcal{M}^{ }_\rmi{f}|^2 
 \;. \nn \la{boltzmann}
\ea 
Here the chemical potentials read
$\muLa = \mu^{ }_a - \muY^{ }/2$, 
$\muH^{ }= \muY^{ }/2$,  
$
 \mutL  = \frac{\muY^{ }}{6} + \frac{\muB^{ }}{3}
$
and
$
 \mutR  = \frac{2\muY^{ }}{3} + \frac{\muB^{ }}{3}
$, and ${\rm d}\Omega^{ }_{n\to m}$ denotes the phase space 
integration measure. Furthermore $p_i \equiv |\vec{p}_i|$
denote incoming and $k_i \equiv |\vec{k}_i|$ outgoing momenta.
The matrix elements read 
\ba
 && \hspace*{-4mm}
 |\mathcal{M}^{ }_\rmi{a}|^2  \; = \; 
 (g_1^2 + 3 g_2^2) 
 \, \frac{u}{t} 
 \;,  \quad 
 |\mathcal{M}^{ }_\rmi{b}|^2 \; = \;
 - (g_1^2 + 3 g_2^2) 
 \, \frac{u}{s} 
 \;, \quad 
 |\mathcal{M}^{ }_\rmi{c}|^2 \; = \;
 - (g_1^2 + 3 g_2^2) 
 \,
   \frac{s}{u} 
 \;, \hspace*{8mm} \la{2z2} \\ 
 && \hspace*{-4mm}
 |\mathcal{M}^{ }_\rmi{d}|^2 \; = \;
 |\mathcal{M}^{ }_\rmi{e}|^2 \; = \; 
 |\mathcal{M}^{ }_\rmi{f}|^2 \; = \; 
 2 h_t^2 \Nc
 \;, \la{M*M}
\ea 
where $s$, $t$ and $u$ are the Mandelstam variables. 

Generalizing the techniques of ref.~\cite{bb2} and parametrizing by
$q^{ }_{\pm} \equiv (q^{ }_0 \pm q)/2$ the 4-momentum of
an exchanged particle, all but two of the phase
space integrals can be carried out, yielding 
\ba
 && \hspace*{-1.5cm}
 \tr \{
  \bsl{\mathcal{K}} 
  \, \aL \,  \rho^{2\leftrightarrow 2,\rmi{hard}}_a(\mathcal{K}) \,  \aR \,
 \} \nn 
 & = & 
 \frac{1}{(4\pi)^3 k }
 \int_{ k }^{\infty} \! {\rm d} \qp
 \int_0^{ k } \! {\rm d} \qm 
 \Bigl\{ 
  \bigl[\nB{}(q^{ }_0 - \muH)
  + \nF{}(q^{ }_0 - k + \muLa) \bigr] 
  \,  \Phi_\rmi{$s$b} 
 \nn & & \quad
 + \, 
  \bigl[\nF{}(q^{ }_0 - \muLa)
  + \nB{}(q^{ }_0 - k + \muH) \bigr] 
  \,  \Phi_\rmi{$s$f} 
 \Bigr\} 
 \nn  
 & + & 
 \frac{1}{(4\pi)^3 k }
 \int_{0}^{ k } \! {\rm d} \qp
 \int_{-\infty}^{0} \! {\rm d} \qm  
 \Bigl\{ 
  \bigl[1 + \nB{}(q^{ }_0 - \muH)
  - \nF{}(k - q^{ }_0 - \muLa) \bigr]
  \, \Phi_\rmi{$t$b} 
 \nn & & \quad
 + \, 
  \bigl[1 - \nF{}(q^{ }_0 - \muLa) + \nB{}(k - q^{ }_0 - \muH) \bigr]
  \, \Phi_\rmi{$t$f} 
 \Bigr\} 
 \;. \la{2to2}
\ea
Introducing the functions 
\ba
 && 
 \lnf(q) \; \equiv \; \ln \Bigl( 1 + e^{-q/T} \Bigr)
 \;, \quad
 \lif(q) \; \equiv \; \mbox{Li}^{ }_2 \Bigl(-e^{-q/T}\Bigr)
 \;, \\
 && 
 \lnb(q) \; \equiv \; \ln \Bigl( 1 - e^{-q/T} \Bigr)
 \;, \quad
 \lib(q) \; \equiv \; \mbox{Li}^{ }_2 \Bigl(e^{-q/T}\Bigr)
 \;, 
\ea
processes with bosonic and fermionic $s$-channel exchange lead to 
\ba
 \Phi_\rmi{$s$b} & = & 
 2 h_t^2 \Nc \, 
 \Bigl\{ 
 q + T \, 
 \Bigl[ 
   \lnf(\qp + \mutL)
 + \lnf(\qp - \mutR)
 \nn 
 & - & \lnf(\qm + \mutL)
 - \lnf(\qm - \mutR)
 \Bigr]
 \Bigr\} 
 \;, \la{Phisb} \\ 
 \Phi_\rmi{$s$f} & = & 
 (g_1^2 + 3 g_2^2) \biggl\{ 
 \frac{q}{2} + \frac{T}{q} 
 \Bigl[ (k-\qm)\Bigl( \lnf(\qp - \muLa) - \lnb(\qm) \Bigr) 
 \nn 
 & + & (k - \qp) \Bigl( \lnf(\qm - \muLa) - \lnb(\qp) \Bigr)  
 \Bigr]
 \nn 
 & + & \frac{(q^{ }_0 - 2k)T^2}{q^2}
 \Bigl[ 
  \lif(\qp - \muLa) - \lif(\qm - \muLa) 
 \; + \; \lib(\qm) - \lib(\qp) 
 \Bigr]\biggr\} 
 \;. \la{Phisf}
\ea
The corresponding $t$-channel contributions read
\ba
 \Phi_\rmi{$t$b} & = & 
 2 h_t^2 \Nc \, 
 T \, \Bigl[ 
   \lnf(-\qm - \mutL)
 + \lnf(-\qm + \mutR)
 \nn 
 & - & \lnf(\qp + \mutL)
 - \lnf(\qp - \mutR)
 \Bigr]
 \;, \la{Phitb} \\ 
 \Phi_\rmi{$t$f} & = & 
 (g_1^2 + 3 g_2^2) \biggl\{ 
 \frac{T}{q} 
 \Bigl[ (k-\qm)\Bigl( \lnf(\qp - \muLa) - \lnb(-\qm) \Bigr) 
 \nn 
 & + & (k - \qp) \Bigl( \lnf(-\qm + \muLa) - \lnb(\qp) \Bigr)  
 \Bigr]
 \nn 
 & + & \frac{(q^{ }_0 - 2k)T^2}{q^2}
 \Bigl[ 
  \lif(\qp - \muLa) + \lif(-\qm + \muLa) 
 \; - \; \lib(-\qm) - \lib(\qp) 
 \Bigr]\biggr\} 
 \;. \hspace*{5mm} \la{Phitf}
\ea

%
\subsection{Resummation of soft $t$-channel lepton exchange}
\la{ss:lepton_exchange}

As already mentioned 
the massless matrix elements and phase space integrals 
lead to logarithmic IR divergences. A well-known divergence originates from 
fermionic $t$-channel exchange around
$(\qp,\qm) \approx (0,0)$ where the integrand can be 
approximated as 
\be
 \Phi_\rmi{$t$f}\; \approx \; (g_1^2 + 3 g_2^2) \frac{ 2 k T^2 }{ q^2 }
 \bigl[
   2\lib(0)  - \lif(- \muLa) - \lif(\muLa) 
 \bigr]
 \; = \; (g_1^2 + 3 g_2^2) \frac{ k  }{ q^2 }
 \bigl[
  \pi^2 T^2 + \muLa^2 
 \bigr]
 \;. \la{IRdiv}
\ee
The divergence is regulated by a 
thermal mass, denoted by $m^{ }_{\ell,a}$, that the lepton obtains
through its interactions with the Standard Model plasma: 
\be
 m_{\ell,a}^2 = \frac{g_1^2 + 3 g_2^2}{16}
 \Bigl( T^2 + \frac{\muLa^2}{\pi^2} \Bigr) 
 \;. \la{mell}
\ee
Computing 
the contribution of soft momenta $q^{ }_\perp \sim m^{ }_{\ell}$
requires a Hard Thermal Loop (HTL) resummed 
computation~\cite{bb2}. Fortunately, this computation remains 
practically identical in the presence of a chemical potential, 
so we just briefly state the results.

Following the presentation in \se{4.1} of ref.~\cite{broken}, 
the resummed computation yields two separate ingredients. One is 
a ``subtraction term'' which removes the IR divergence
in \eq\nr{IRdiv} from the naive computation:
\ba
 && \hspace*{-1.5cm} 
 \tr \{
  \bsl{\mathcal{K}} 
  \, \aL \, 
 \rho^{2\leftrightarrow 2,\rmi{subtr}}_a(\mathcal{K}) \,  \aR \,
 \} 
 \nn[2mm] 
 & = & 
 \frac{1}{(4\pi)^3}
 \int_{0}^{k} \! {\rm d} \qp \int_{-\infty}^{0} \! {\rm d} \qm  
 \, \bigl[\nB(k - \muH^{ }) + \nF(\muLa) \bigr]
 \, (g_1^2 + 3g_2^2) \,  \frac{\pi^2 T^2 + \muLa^2}{q^2}
 \;. 
\ea
The second ingredient 
is the correctly computed IR contribution. For this we obtain
\be
  \tr \{
  \bsl{\mathcal{K}} 
  \, \aL \,  \rho^{2\leftrightarrow 2,\rmi{soft}}_a(\mathcal{K}) \,  \aR \,
 \} 
 \; = \; 
 \frac{m_{\ell,a}^2}{8\pi}
 \ln\Bigl[ 1 + \Bigl(\frac{2 k}{m^{ }_{\ell,a}} \Bigr)^2 \Bigr]
 \, \bigl[\nB(k - \muH^{ }) + \nF(\muLa) \bigr]
 \; + \; 
 \rmO\Bigl( \frac{m_{\ell,a}^4}{k^2} \Bigr)
 \;. 
\ee
In total, the $2\leftrightarrow 2$
contribution can then be expressed as 
$ 
 \tr \{
  \bsl{\mathcal{K}} 
  \, \aL \,  \rho^{2\leftrightarrow 2}_a(\mathcal{K}) \,  \aR \,
 \}  
$
where 
\be 
 \rho^{2\leftrightarrow 2}_a(\mathcal{K}) 
 \; = \;  
 \rho^{2\leftrightarrow 2,\rmi{hard}}_a(\mathcal{K}) 
 - 
 \rho^{2\leftrightarrow 2,\rmi{subtr}}_a(\mathcal{K})
 + 
 \rho^{2\leftrightarrow 2,\rmi{soft}}_a(\mathcal{K})
 \;. 
\ee

%
\subsection{Resummation of soft $s$ and $t$-channel Higgs scattering}
\la{ss:external_Higgs}

There is another IR divergence which at leading order 
only affects chemical potential
dependence, specifically $S^{ }_{(\tau)}$ 
defined in accordance with \eq\nr{QRS1}. 
It originates from the fact that when expanded in $\muH = \muY/2$, 
the bosonic distribution functions multiplying 
$ \Phi^{ }_\rmi{$s$f} $ and 
$ \Phi^{ }_\rmi{$t$f} $ in \eq\nr{2to2} diverge as 
$\sim \pm \muH T /(k-q^{ }_0)^2$ at one corner of the integration range, 
cf.\ \fig\ref{fig:phasespace}. 
When the integration is defined as a principal value, most terms
cancel between $s$ and $t$-channel contributions. 
However, a small remainder is left over. 

To be specific, 
the distribution functions appearing in the $2\leftrightarrow 2$ contributions
can be expanded as in \eqs\nr{Q12}-\nr{S12}, 
whereas \eqs\nr{Phisb}--\nr{Phitf} read
\ba
 \Phi^{ }_\rmi{$s$b}(\{\mu_i\}) & = & 
 \Phi^{ }_\rmi{$s$b}(0) + \bmuY
  h_t^2 \Nc T  \bigl[ \nF(\qp) - \nF(\qm) \bigr] + \rmO(\bar{\mu}^2)
 \;, \\ 
 \Phi^{ }_\rmi{$s$f}(\{\mu_i\}) & = & 
 \Phi^{ }_\rmi{$s$f}(0) + 
 \Bigl(\bar{\mu}^{ }_a - \frac{\bmuY}{2} \Bigr)
 (g_1^2 + 3 g_2^2)
 \biggl\{ 
 \frac{T}{q} 
 \bigl[
  ( k - \qm) \nF(\qp) + (k - \qp) \nF(\qm)  
 \bigr]
 \nn 
 & + & \frac{(q^{ }_0 - 2 k)T^2}{q^2} 
 \bigl[ \lnf(\qm) - \lnf(\qp) \bigr] \biggr\} + \rmO(\bar{\mu}^2)
 \;, \\ 
 \Phi^{ }_\rmi{$t$b}(\{\mu_i\}) & = & 
 \Phi^{ }_\rmi{$t$b}(0) - \bmuY
  h_t^2 \Nc T  \bigl[ \nF(\qp) + \nF(-\qm) \bigr] + \rmO(\bar{\mu}^2)
 \;, \\ 
 \Phi^{ }_\rmi{$t$f}(\{\mu_i\}) & = & 
 \Phi^{ }_\rmi{$t$f}(0) + 
 \Bigl(\bar{\mu}^{ }_a - \frac{\bmuY}{2} \Bigr)
 (g_1^2 + 3 g_2^2)
 \biggl\{ 
 \frac{T}{q} 
 \bigl[
  ( k - \qm) \nF(\qp) - (k - \qp) \nF(-\qm)  
 \bigr]
 \nn 
 & + & \frac{(q^{ }_0 - 2 k)T^2}{q^2} 
 \bigl[ \lnf(-\qm) - \lnf(\qp) \bigr] \biggr\} + \rmO(\bar{\mu}^2)
 \;. 
\ea
The problem originates from the fact that 
$\Phi^{ }_\rmi{$s$f}(0)$ and
$\Phi^{ }_\rmi{$t$f}(0)$ are not equal 
when approaching $(q,q^{ }_0) = (k,k)$ from the $s$ and $t$-channel sides, 
respectively. 

In order to cure the problem, 
we may subtract the divergent terms from the integrand of 
\eq\nr{2to2} (denoting by $\Delta$ terms within the curly brackets), 
\ba
 \Delta^{ }_\rmi{$s$f} & \equiv & 
 - \frac{(g_1^2 + 3 g_2^2) \muH T}{(k-q^{ }_0)^2}
 \biggl[ \frac{k}{2} - T \ln\biggl(\frac{\qm}{T}\biggr) - 
        \frac{\pi^2T^2}{2k} + \chi(k)  \biggr]\;, 
 \quad k < q^{ }_0 < 2k 
 \;,  \hspace*{8mm} \\ 
 \Delta^{ }_\rmi{$t$f} & \equiv & 
 \frac{(g_1^2 + 3 g_2^2) \muH T}{(k-q^{ }_0)^2}
 \biggl[ - T \ln\biggl(-\frac{\qm}{T}\biggr) + \chi(k)  \biggr]\;, 
 \quad
  0 < q^{ }_0 < k 
 \;, 
\ea
where the function $\chi$ reads
$
 \chi(k) = T \lnf(k) + \frac{T^2}{k}[\frac{\pi^2}{4} + \lib(k) - \lif(k) ]
$. 
Subsequently, the $s$-channel subtraction is reflected into 
the $t$-channel domain by $q^{ }_0 \to 2k - q^{ }_0$, $q\to 2k - q$, 
whereby the logarithms and $\chi$ drop out. 
The remainder is integrated after noting that the $t$-channel
integration domain in \fig\ref{fig:phasespace}
originates from the energy-conservation constraint 
$\delta(q^{ }_0 - k + \epsilon^{\phi}_{\vec{k-q}})$, 
where a soft (i.e.\ $|\vec{k-q}| \lsim gT$) Higgs energy is 
$\epsilon^{\phi}_{\vec{k-q}} \equiv \sqrt{(\vec{k-q})^2+m_\phi^2}$. 
Working out the $t$-channel integration range in the presence
of $m^{ }_\phi > 0$ yields the correct IR contribution from 
soft Higgs scattering to $S^{2\leftrightarrow 2}_{(+)}$, 
\ba
 \Delta S^{2\leftrightarrow 2}_{(+)} & = & 
 \frac{g_1^2 + 3 g_2^2}{(4\pi)^3 4k}
 \int_0^{k-m^{ }_\phi} \! {\rm d}q^{ }_0 
 \int_{k - \sqrt{(k - q^{ }_0)^2 - m_\phi^2}}
     ^{k + \sqrt{(k - q^{ }_0)^2 - m_\phi^2}} \! {\rm d}q \, 
 \frac{-T}{(k - q^{ }_0)^2}
 \biggl[
  \frac{k}{2} - \frac{\pi^2T^2}{2k}
 \biggr]
 \nn 
 & = & 
 \frac{(g_1^2 + 3 g_2^2)T}{4(4\pi)^3}
 \biggl( \frac{\pi^2 T^2}{k^2} - 1 \biggr)
 \biggl[ \ln\biggl( \frac{2 k}{m^{ }_\phi} \biggr) - 1 \biggr]
 + \rmO\biggl( \frac{m^{ }_\phi T }{k} \biggr)
 \;. 
\ea

%
\subsection{Numerical results}
\la{ss:numerics}

\begin{figure}[t]

\hspace*{-0.1cm}
\centerline{%
 \epsfxsize=7.5cm\epsfbox{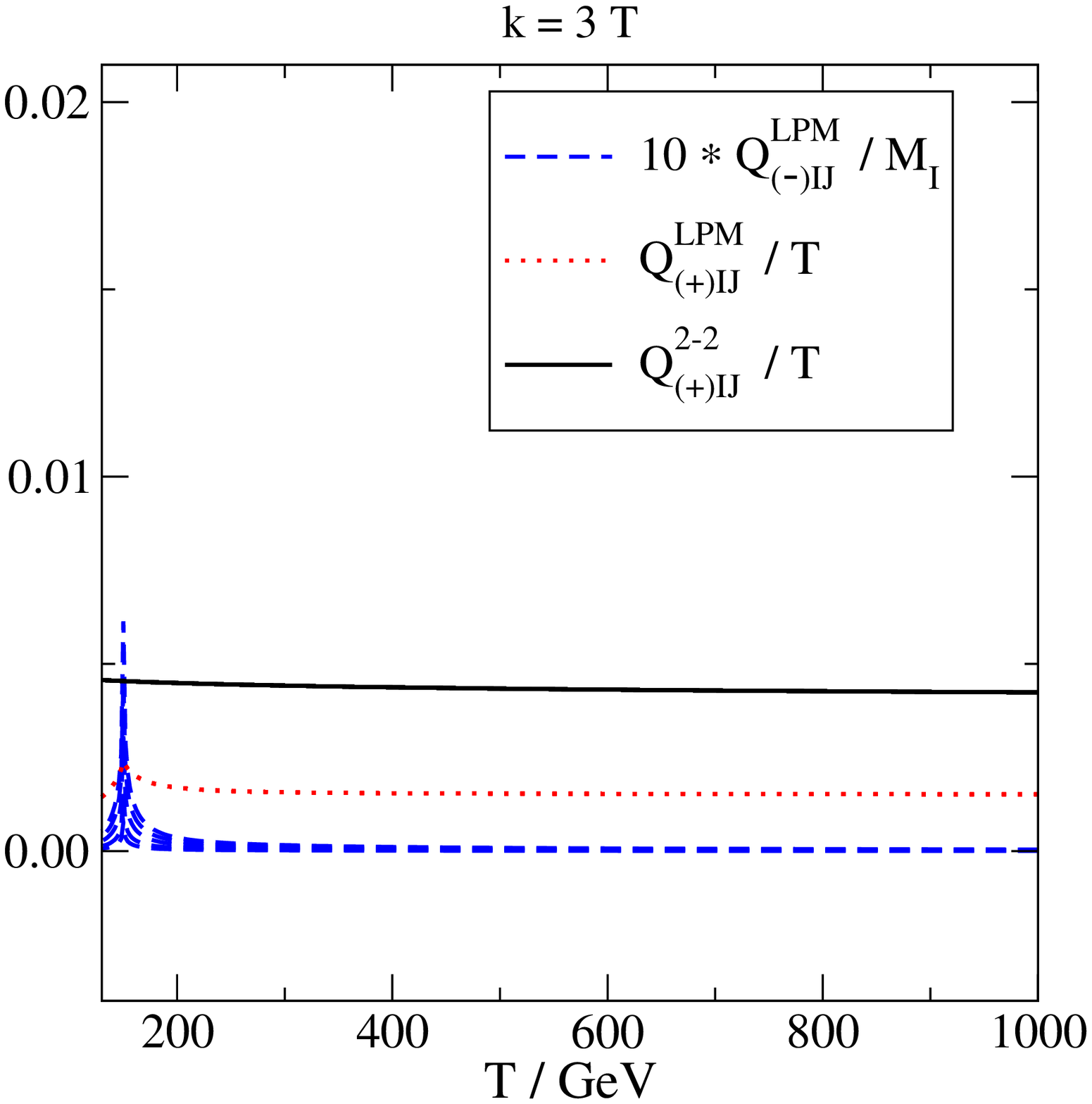}%
 \hspace{0.1cm}
 \epsfxsize=7.5cm\epsfbox{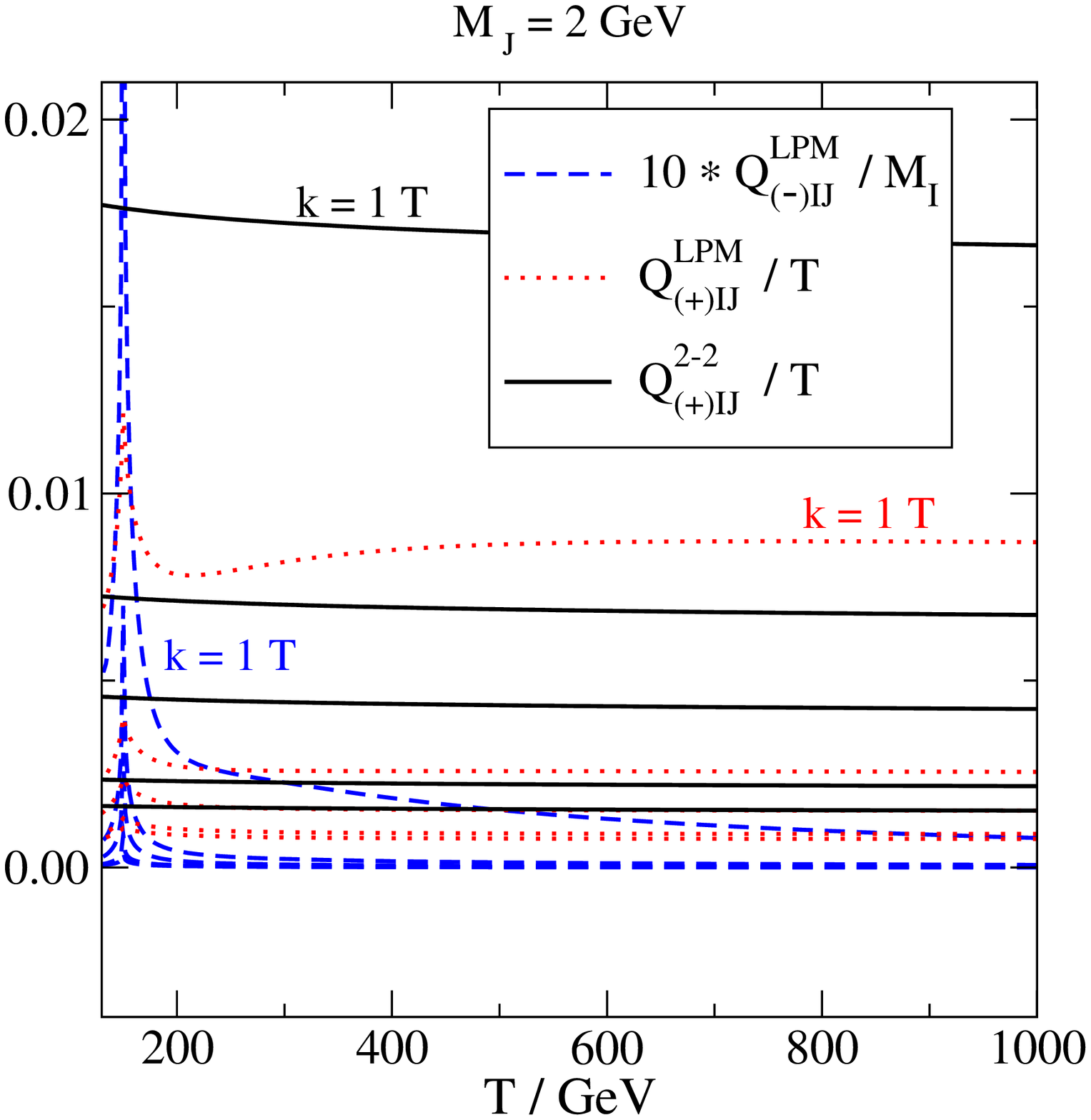}
}

\caption[a]{\small
 Left: The coefficient $Q^{ }_{(\tau)\I\J}$ from
 \eq\nr{QRS1}, for fixed $k=3T$ and $M^{ }_{\J}$/GeV $\in \{0.5,1,2,3,4\}$; 
 the dependence on $M^{ }_{\J}$ is moderate, and the dependence on $M^{ }_{\I}$
 is exactly cancelled by the normalization chosen. There is a mild divergence
 at the location of the electroweak crossover, indicating that the 
 perturbative computation becomes unreliable there.   
 Right: The same for fixed $M^{ }_{\J}=2$~GeV and $k/T \in \{1,2,3,6,9\}$. 
 One curve has been labelled, with the dependence on $k/T$ being monotonic. 
}

\la{fig:coeffQ}
\end{figure}

\begin{figure}[t]

\hspace*{-0.1cm}
\centerline{%
 \epsfxsize=7.5cm\epsfbox{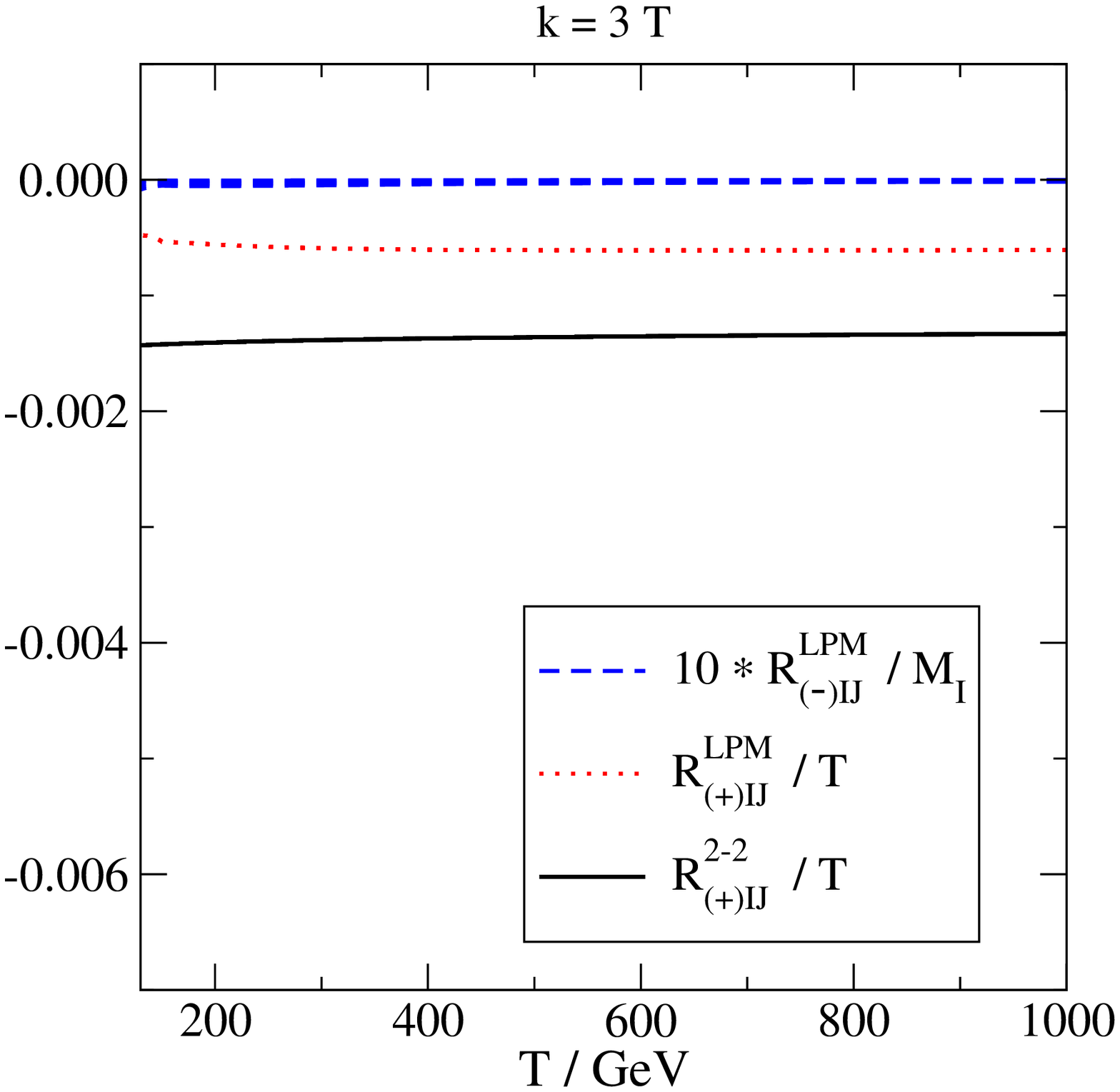}%
 \hspace{0.1cm}
 \epsfxsize=7.5cm\epsfbox{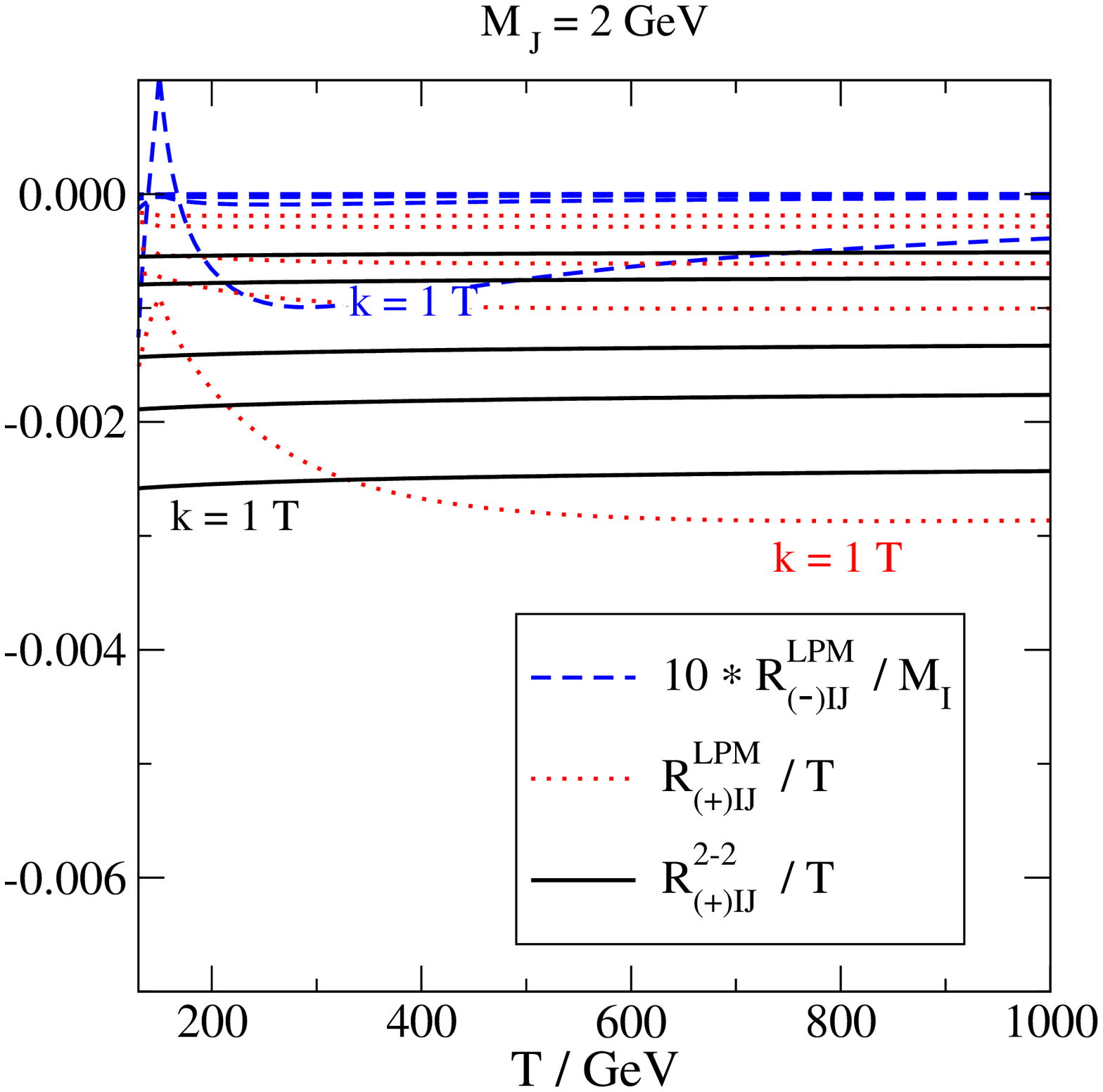}
}

\caption[a]{\small
 Left: $R^{ }_{(\tau)\I\J}$ from
 \eq\nr{QRS1}, for fixed $k=3T$ and $M^{ }_{\J}$/GeV $\in \{0.5,1,2,3,4\}$. 
 Mild mass dependence is seen in $R^{ }_{(-)\I\J}$, 
 but $R^{ }_{(-)\I\J}$ is very small once multiplied 
 by $M^{ }_{\I}/T\sim 10^{-2}$ in order to express it in the same
 units as the other contributions. 
 Right: The same for fixed $M^{ }_{\J}=2$~GeV and $k/T \in \{1,2,3,6,9\}$. 
 One curve has been labelled, with the dependence on $k/T$ being monotonic. 
}

\la{fig:coeffR}
\end{figure}

\begin{figure}[t]

\hspace*{-0.1cm}
\centerline{%
 \epsfxsize=7.5cm\epsfbox{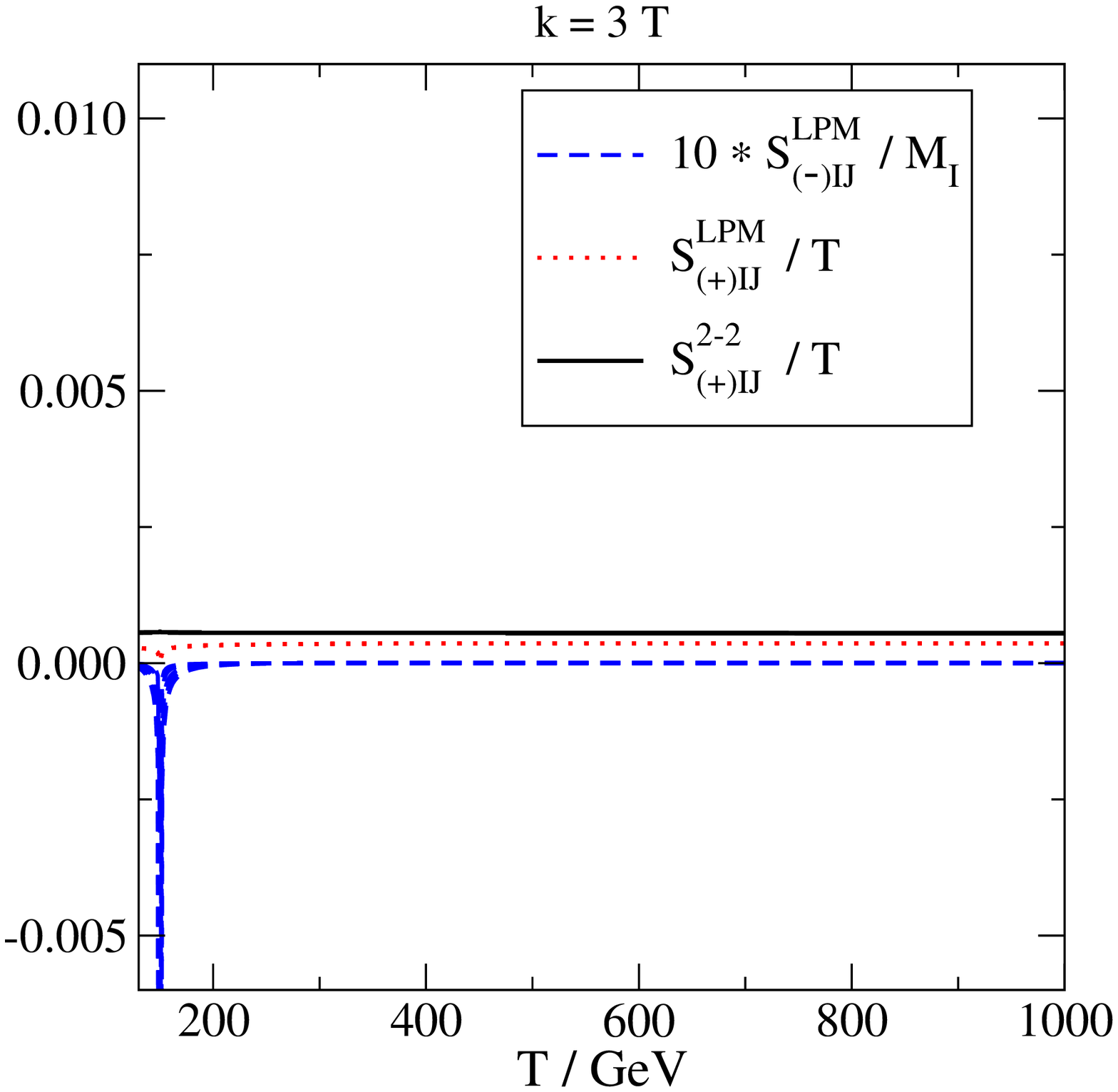}%
 \hspace{0.1cm}
 \epsfxsize=7.5cm\epsfbox{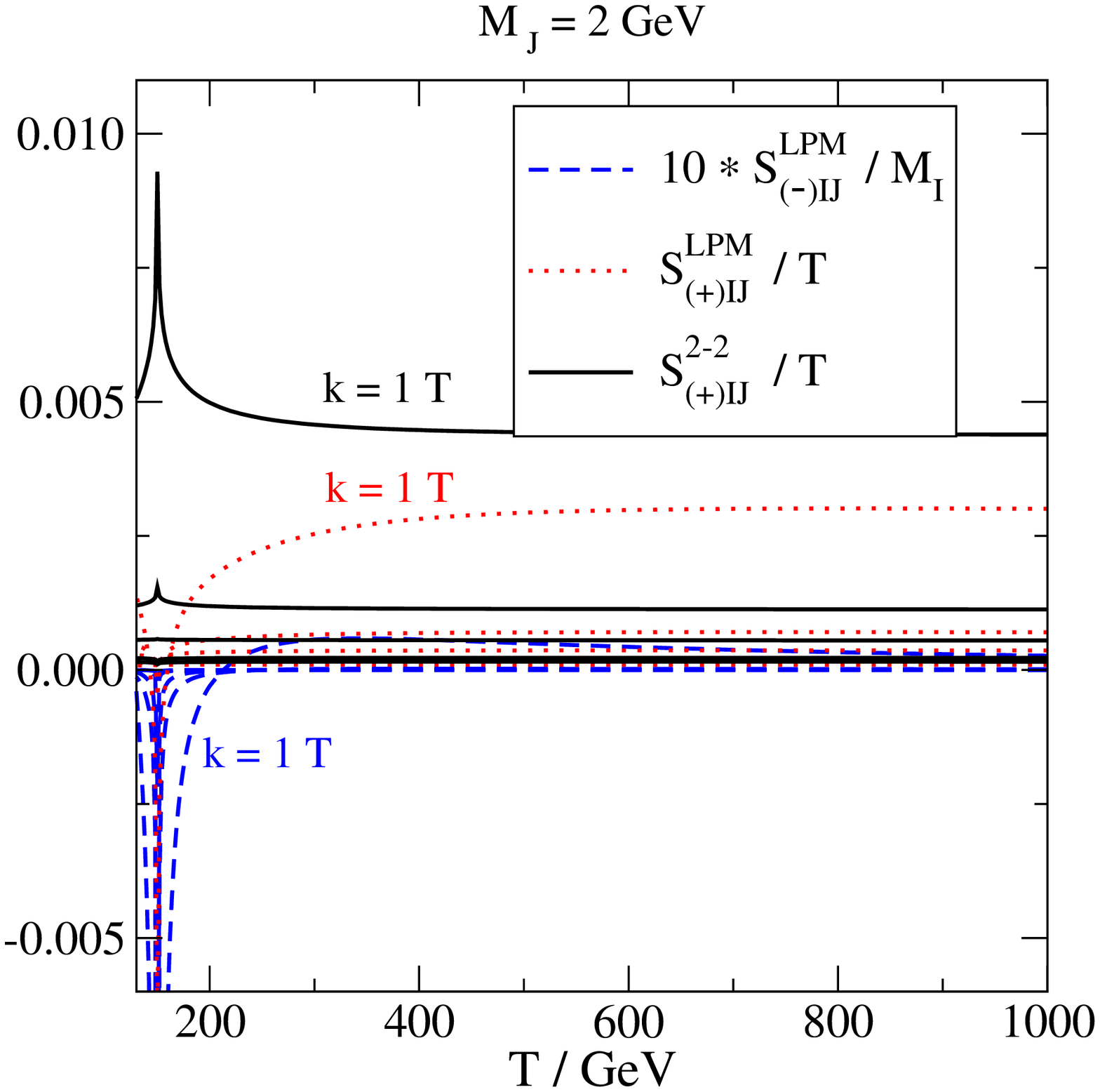}
}

\caption[a]{\small
 Left: $S^{ }_{(\tau)\I\J}$ from
 \eq\nr{QRS1}, for fixed $k=3T$ and $M^{ }_{\J}$/GeV $\in \{0.5,1,2,3,4\}$;  
 the mass dependence is mild. 
 Right: The same for fixed $M^{ }_{\J}=2$~GeV and $k/T \in \{1,2,3,6,9\}$. 
 One curve has been labelled, with the dependence on $k/T$ being monotonic. 
}

\la{fig:coeffS}
\end{figure}

We have evaluated the coefficients 
$Q^{ }_{(\tau)}$, $R^{ }_{(\tau)}$, $S^{ }_{(\tau)}$ 
defined in \eqs\nr{QRS1} and \nr{QRS2} numerically
for $T \gsim 130$~GeV.\footnote{%
 The coefficients are well-defined also at $T < 130$~GeV and could  
 be evaluated following ref.~\cite{broken}. 
 } 
We display separately
the $1\leftrightarrow 2$ contributions from 
\eqs\nr{rho_LPM}--\nr{S12} and the 
$2\leftrightarrow 2$ contributions 
from \ses\ref{ss:22}--\ref{ss:external_Higgs}. 
Results for $Q^{ }_{(\tau)}$ are shown 
in \fig\ref{fig:coeffQ}, those for $R^{ }_{(\tau)}$ 
in \fig\ref{fig:coeffR}, and those for $S^{ }_{(\tau)}$ 
in \fig\ref{fig:coeffS}.

We find that the helicity-flipping
coefficients 
$Q^{ }_\rmii{(+)} = Q^\rmii{LPM}_\rmii{(+)} 
+ Q^\rmii{$2\!\leftrightarrow\! 2$}_\rmii{(+)}$, 
$|R^{ }_\rmii{(+)}|$, and 
$S^{ }_\rmii{(+)}$
are of order 
$\sim (10^{-3} ... 10^{-2}) T$, with in general 
negative values for $R^{ }_\rmii{(+)}$.
The helicity-conserving coefficients
$Q^{ }_\rmii{($-$)}$, 
$R^{ }_\rmii{($-$)}$, and 
$S^{ }_\rmii{($-$)}$ are suppressed by sterile neutrino masses, because in 
the massless limit right-handed neutrinos carry opposite
helicity to left-handed Standard Model leptons.
In the figures these coefficients 
were normalized to $M^{ }_{\I}$; when normalized
to $T$, their 
contribution is suppressed by $M^{ }_{\I} / T\sim 10^{-2}$. Therefore
the coefficients appearing in \eqs\nr{A}--\nr{Bminus}, \nr{Cplus}--\nr{Dminus}
are dominated by the helicity-flipping contributions.
However the helicity-conserving coefficients are more
IR sensitive than the helicity-flipping ones, 
showing a mild divergence
around the crossover at which their perturbative determination becomes
unreliable, and they also dictate the fermion number violation
rate in accordance with \se\ref{ss:fermion}, cf.\ \eqs\nr{E}--\nr{G}.

%
\section{Relation of chemical potentials and lepton asymmetries}
\la{se:mu}

%
\subsection{General setup}
\la{ss:general_mu}

The left-hand side of \eq\nr{summary_na}
contains charge densities, whereas on
the right-hand sides of \eqs\nr{summary_na}, \nr{summary_rho_plus} 
and \nr{summary_rho_minus} 
chemical potentials appear. 
In order to close the set of equations, the chemical potentials need to 
expressed in terms of the charge densities. To leading order the 
results are given by \eqs\nr{mu_highT} and \nr{mu_lowT},
whose derivations we wish to briefly review.  

Charge neutrality of the plasma poses a non-trivial constraint
on the relation between chemical potentials and number densities.
In the temperature range of interest we can
to a good approximation assume the electroweak symmetry to be restored.
Then charge neutrality concerns the hypercharge field. Defining 
a corresponding chemical potential as 
$
 \muY^{ } \equiv i g^{ }_1 {B}^{ }_0 
$, 
where $B^{ }_0$ is the hypercharge field in the imaginary-time formalism, 
a simple way to proceed is to first express the pressure
(minus the free energy density) in terms of $\muY^{ },\mu^{ }_a$
and $\muB^{ }$~\cite{khlebnikov}. To leading order in Standard Model 
couplings, treating all particles as massless
(masses are included in \eq(A.6) of ref.~\cite{broken}), 
we obtain 
\ba
 p(T,\mu) - p(T, 0 ) & = & 
 \Bigl[ 2 \muB^2 + 2 \muY^{ } \muB^{ } +
 \sum_a \Bigl( \fr32  \mu_a^2 
 -2 \muY^{ } \mu^{ }_a \Bigr) 
 + 5 \muY^2 \Bigr]\, \chi^{ }_\rmii{F}(0)
 \nn 
 & + &  \frac{\muY^2}{4} \chi^{ }_\rmii{B}(0)
 + \rmO(g\mu^2, \mu^4) \;, \la{pmu}
\ea
where 
$
 \chi^{ }_\rmii{F}(m) \equiv  
 \int_\vec{k} \bigl[ - 2 \nF'(\omega^k) \bigr] 
$ 
and
$
 \chi^{ }_\rmii{B}(m)  \equiv  
 \int_\vec{k} \bigl[ - 2 \nB'(\omega^k) \bigr] 
$
are so-called susceptibilities, 
with the special values 
$
 \chi^{ }_\rmii{F}(0) = T^2/6
$,
$
 \chi^{ }_\rmii{B}(0) = T^2/3
$.
Hypercharge neutrality corresponds now to 
$\partial p / \partial \muY^{ }  = 0$, 
and the conserved charge densities are obtained as 
$\partial p / \partial \mu^{ }_a$, $\partial p / \partial \muB^{ }$. 

%
\subsection{$T > 130$~GeV}
\la{ss:highT}

At $T > 130$~GeV the baryon chemical potential 
is eliminated through the sphaleron constraint 
$\muB^{ } = -\fr13 \sum_a \mu^{ }_a $, so that $\mu^{ }_a$
couples to the strictly conserved quantity $L^{ }_a - \fr13 B$. 
In a path integral formalism, the presence of $\mu^{ }_a \neq 0$
implies that the perturbative minimum lies at a non-zero value
of $\muY$, as determined by \eq\nr{pmu}. Minimizing
with respect to $\muY^{ }$ we obtain
\be
 \muY^{ } = \frac{8}{33} \sum_a \mu^{ }_a  + \rmO(g)
 \;, \la{muY_high}
\ee
and 
\be
 \left( 
  \begin{array}{c}
    \mu^{ }_1 \\
    \mu^{ }_2 \\ 
    \mu^{ }_3    
  \end{array}
 \right)
 \; = \; \frac{1}{237 T^2} 
 \left( 
   \begin{array}{ccc} 
     514 & 40 & 40 \\  
     40 & 514 & 40 \\ 
     40 & 40 & 514 
   \end{array}
 \right)
 \left( 
  \begin{array}{c}
    n^{ }_1 - \frac{\nB}{3} \\
    n^{ }_2 - \frac{\nB}{3} \\ 
    n^{ }_3 - \frac{\nB}{3}  
  \end{array} 
 \right) + \rmO(g)
 \;. \la{mu_highT}
\ee
The numerical uncertainties of these expression are about 20\%, 
owing mostly to large $O(\alphas)$ corrections from 
the QCD coupling~\cite{washout} 
and to IR sensitive effects from the Higgs~\cite{susc}. 
The baryon asymmetry is given by~\cite{khleb0}
\be
 \nB = \frac{\partial p}{\partial \muB^{ }} = 
 \bigl[ 4 \muB^{ }+ 2 \muY^{ } \bigr] \chi^{ }_\rmii{F}(0)
 = -\frac{14 T^2}{99}\sum_a \mu^{ }_a = 
 -\frac{28}{79} \sum_a \Bigl( n^{ }_a - \frac{\nB}{3} \Bigr)
 \;. \la{nB_highT} 
\ee

%
\subsection{$T \sim 130$~GeV}
\la{ss:lowT}

When $T \sim 130$~GeV, the sphaleron processes become slow. Consequently, 
baryon plus lepton asymmetry needs to be added as a dynamical non-equilibrium 
variable. The quantities for which equations of motion can be written
are $n^{ }_a - \nB^{ }/3$ and $\nB^{ } + \sum_a n^{ }_a$. Coupling
chemical potentials to these slow variables we can read off the
original chemical potentials $\mu^{ }_a$ and $\muB^{ }$: 
\be
 \sum_a \tilde{\mu}^{ }_a \Bigl( n^{ }_a - \frac{\nB^{ }}{3} \Bigr) 
 + \tmuBL \bigl( \nB^{ } + \sum_a n^{ }_a \bigr)
 \; = \; 
 \sum_a \underbrace{\bigl( \tilde{\mu}^{ }_a + \tmuBL \bigr)}
  _{\mu^{ }_a} \, n^{ }_a
 + 
 \underbrace{\Bigl(\tmuBL - \frac{\sum_a \tilde{\mu}^{ }_a}{3} \Bigr)}
  _{\muB^{ }} \nB^{ }
 \;. \la{muY_low}
\ee
These values of $\mu^{ }_a$ and $\muB^{ }$ are inserted into
\eq\nr{pmu}. The pressure is minimized with respect to $\muY^{ }$
like before, leading to
\ba
 \muY^{ } \; = \; \frac{8}{33} \Bigl( \sum_a \tilde{\mu}^{ }_a +
 \frac{ 3 \tmuBL^{ }}{2} \Bigr) 
 + \rmO(g) 
 \;. \hspace*{8mm} \la{mut_low}
\ea
Furthermore, taking partial derivatives of 
\eq\nr{pmu} with respect to $\tilde{\mu}^{ }_a$ and $\tmuBL$, 
we obtain $n^{ }_a - \nB^{ }/3$ 
and $\nB^{ }+ \sum_a n^{ }_a$ as functions of the chemical 
potentials. Inverting these relations, we get
\be
 \left( 
  \begin{array}{c}
    \tilde{\mu}^{ }_1 \\
    \tilde{\mu}^{ }_2 \\ 
    \tilde{\mu}^{ }_3 \\
    \tmuBL^{ }   
  \end{array}
 \right)
 \; = \; \frac{1}{144 T^2} 
 \left( 
   \begin{array}{rrrr} 
     319 & 31 & 31 & -23 \\  
     31 & 319 & 31 & -23 \\ 
     31 & 31 & 319 & -23 \\
     -23 & -23 & -23 & 79
   \end{array}
 \right)
 \left( 
  \begin{array}{c}
    n^{ }_1 - \frac{\nB^{ }}{3}  \\
    n^{ }_2 - \frac{\nB^{ }}{3} \\ 
    n^{ }_3 - \frac{\nB^{ }}{3} \\
   \nB + \sum_a n^{ }_a  
  \end{array} 
 \right) + \rmO(g)
 \;. \la{mu_lowT}
\ee
Equations \nr{muY_low}--\nr{mu_lowT} fix the right-hand sides of 
\eqs\nr{summary_na}, \nr{summary_rho_plus} and \nr{summary_rho_minus} 
in terms of the slowly 
evolving number densities. As a crosscheck, if we fix 
$\nB^{ }$ from \eq\nr{nB_highT}, 
\eq\nr{mu_lowT} yields $\tmuBL = 0$; according
to \eq\nr{summary_nB} this indeed corresponds to
a stationary state.

%
\section{Evolution of baryon plus lepton asymmetry}
\la{se:baryon}

Suppose that we start the evolution of the system 
from a high temperature, $T \gg 130$~GeV, and
are given some initial values of the lepton symmetries $n^{ }_a - \nB^{ }/3$, 
for instance $n^{ }_a - \nB^{ }/3 = 0 \; \forall a$. 
To solve the evolution equations \nr{summary_na}, \nr{summary_rho_plus}
and \nr{summary_rho_minus}, 
we first need to determine the chemical potentials $\mu^{ }_a$. 
These can be obtained from \eq\nr{mu_highT}. 
The baryon asymmetry is known 
as a ``side product'' of the evolution,
from \eq\nr{nB_highT}.  

The situation changes when the sphaleron processes become slow. 
We can switch to this setting at some temperature $T^{ }_0 > 130$~GeV
at which we know the initial values of $n^{ }_a - \nB/3$ and 
$\nB^{ }+ \sum_a n^{ }_a$ from
the computation described above
(note that $\nB^{ }+ \sum_a n^{ }_a$ is in general non-zero, and can be
determined from \eq\nr{nB_highT}). 
The corresponding
chemical potentials
can be determined from \eq\nr{mu_lowT}. The other chemical 
potentials are obtained from 
\eqs\nr{muY_low} and \nr{mut_low}, and can then be 
inserted into \eqs\nr{summary_na}, \nr{summary_rho_plus} and
\nr{summary_rho_minus}. 

Obtaining the evolution equation for $\nB^{ }+ \sum_a n^{ }_a$
is non-trivial, given that the fluctuations of  
$n^{ }_a - \nB/3$ and $\nB^{ }+ \sum_a n^{ }_a$ are correlated, 
as exemplified by \eq\nr{mu_lowT}. However the starting point
is again an operator equation of motion analogous to  
\eqs\nr{eom_a} and \nr{eom_L}. This time it takes the form
of the anomaly equation, 
\be
 \dot{B} + \sum_a \dot{L}^{ }_a
  = 2 \nG^{ }\int_\vec{x} c(t,\vec{x}) 
 \;, \la{anomaly}
\ee 
where we have introduced $\nG = 3$ as the number of Standard Model
generations; the factor 2 accounts for baryons and leptons;
and $c$ is the topological charge density. In principle this 
operator could be inserted into \eq\nr{eom_general}, but
it is not easy to express the first order time evolution
of the density matrix in a useful way~\cite{khleb0}. However, 
we can assume that to leading order in $\tmuBL$ and $\tilde{\mu}^{ }_a$, 
the topological charge density is only correlated with itself. 
Moreover, a general argument concerning correlated 
fluctuations~\cite{washout} shows that we can 
write  
\be
 \dot{X}^{ }_a = -
 \frac{1}{2VT}
 \int_{-\infty}^{\infty} \! {\rm d}t \, 
 \Bigl\langle \fr12 \{ \dot{X}^{ }_a(t) , \dot{X}^{ }_c(0) \} \Bigr\rangle
  \, \Xi^{-1}_{cb} \, X^{ }_b
 \;, \la{master}
\ee
where $X^{ }_a$ are general slowly evolving charges. By $\Xi$ we
have denoted a susceptibility matrix; 
its inverse multiplied by
the charges yields the corresponding chemical potential. 
Specifically, $\Xi^{ }_{cb} = 
 \partial^2 p / \partial {\tilde{\mu}^{ }_c} \partial {\tilde{\mu}^{ }_b}$
and $\Xi^{-1}_{(\rmii{$B$+$L$})b} (X^{ }_b/V) = \tmuBL$.

It remains to compute the symmetric correlator
in \eq\nr{master} for the operator on the right-hand side
of \eq\nr{anomaly}. We denote $\int_\vec{x} c(t,\vec{x}) \equiv 
\dot{N}^{ }_\rmii{CS}(t)$ where in the classical limit
$N^{ }_\rmii{CS}$ is the Chern-Simons number. 
It is conventional to shift the time interval to 
run between zero and positive times; making use 
of the time-reversal symmetry of the anticommutator, 
and arguing furthermore that the 
dynamics is dominated by classical configurations which show 
linearly growing diffusive behaviour at large times,
we can write\footnote{%
 Somewhat more precisely, 
 $
 \lim_{t\to\infty} 
 \int_0^t \! {\rm d}t' \, 
 \langle \{ \dot{N}^{ }_\rmii{CS}(t') , 
 \dot{N}^{ }_\rmii{CS}(0) \} \rangle 
 = 
 \lim_{t\to\infty} \frac{{\rm d}}{{\rm d}t}
 \int_0^t \! {\rm d}t'  
 \int_0^t \! {\rm d}t'' \, \langle 
 \dot{N}^{ }_\rmii{CS}(t')  
 \dot{N}^{ }_\rmii{CS}(t'')  \rangle 
 = 
 \lim_{t\to\infty} \frac{{\rm d}}{{\rm d}t}
 \langle
  [N^{ }_\rmii{CS}(t) - N^{ }_\rmii{CS}(0)]^2  
 \rangle
 $. 
} 
\ba
 \frac{1}{V} \int_{-\infty}^{\infty} \! {\rm d}t \, 
 \Bigl\langle \fr12 \{ \dot{N}^{ }_\rmii{CS}(t) , 
 \dot{N}^{ }_\rmii{CS}(0) \} \Bigr\rangle 
 & \simeq & 
 \lim_{t\to \infty} \frac{1}{V t}
 \bigl\langle [ {N}^{ }_\rmii{CS}(t) -  
 {N}^{ }_\rmii{CS}(0) ]^2  \bigr\rangle 
 \; \equiv \; \Gamma^{ }_\rmi{diff}
 \;. \la{sphaleron}
\ea
Here the infinite-volume limit is implicitly understood.
The quantity in \eq\nr{sphaleron} is precisely the one 
estimated with classical lattice gauge theory simulations
in ref.~\cite{sphaleron}.

To summarize, recalling the factor $2\nG$ from \eq\nr{anomaly}
and the factor $1/(2T)$ from \eq\nr{master}, 
the evolution equation for baryon plus lepton
asymmetry obtains the simple form
\be
 \Bigl\langle \dot{n}^{ }_\rmii{B} + \sum_a \dot{n}^{ }_a  \Bigr\rangle
 \; = \;
 - 2 \nG^2\, \Gamma^{ }_\rmi{diff}(T) \, \frac{\tmuBL }{ T}
 + \rmO(\mu^2) \;. \la{summary_nB}
\ee
Here $\tmuBL$ is a linear combination of all slowly evolving 
charges, as given by \eq\nr{mu_lowT}.

%
\section{Summary and outlook}
\la{se:concl}

In this paper we have presented a ``field-theoretic'' derivation of
evolution equations of a coupled system consisting of lepton asymmetries, 
the baryon asymmetry, and a sterile neutrino density matrix. The basic
equations are \nr{summary_na}, \nr{summary_rho_plus},
\nr{summary_rho_minus}, and 
\nr{summary_nB}. Numerical values of the coefficients parametrizing
these equations can be found in \se\ref{ss:numerics} and
in ref.~\cite{sphaleron}. On the right-hand
sides of the equations, 
various chemical potentials appear; to close the system, 
the chemical potentials need to be expressed in terms of the lepton 
and baryon asymmetries, which can be achieved as
re-iterated in \ses\ref{ss:highT} and \ref{ss:lowT}. 

Prior to our work, many studies have appeared in which similar
evolution equations have been derived 
(cf.\ e.g.\ refs.~\cite{singlet,gagnon,shintaro,dg,canetti,
shuve,hk,abada,val,teresi,n2,n3} and references therein). 
The main novelties of our investigation are 
the full inclusion of both helicity-flipping and conserving contributions 
(or, in the language of \se\ref{ss:fermion}, 
fermion-number conserving and violating effects); 
the inclusion of all chemical potentials and 
gauge field expectation values induced by them;
a consistent leading-order computation of all coefficients 
parametrizing the equations, both in the ``symmetric''
and in the ``Higgs'' phase; as well as a formulation
general enough to permit for the treatment of the regime
in which the sphaleron processes gradually switch off. 
We have also gone beyond linear response theory in the treatment
of the sterile neutrino density matrix, permitting for both its 
small and large deviations from equilibrium. Even though we do not
expect any of these improvements to change the previous results by
orders of magnitude, many of them may play a role
if a numerical precision at or below the 20\% level is desired. 

A numerical solution of the evolution equations within the background
of an expanding universe, with 
a
sphaleron 
rate~\cite{sphaleron}
and 
equation of 
state~\cite{crossover,dono}  
inserted from lattice studies,  
poses a non-trivial technical challenge, to
which we hope to return in the near future. 

For some qualitative insight, consider the coefficients 
producing or equilibrating sterile neutrinos, \eqs\nr{Dplus} and \nr{Dminus}. 
At leading order in chemical potentials, 
the coefficients are determined by $Q^{+}_{ }$ and  
$Q^{-}_{ }$, respectively.
Here $Q^{+}_{ }$ contains a sum over helicity-flipping and
conserving contributions, and $Q^{-}_{ }$ their difference, 
cf.\ \eq\nr{symm}. The part parametrized by $Q^{+}_{ }$ generates
a helicity-symmetric density matrix, and $Q^{-}_{ }$ 
a helicity asymmetry. 
Both yield a parametrically similar
contribution to lepton asymmetry, cf.\ \eqs\nr{summary_na}, 
\nr{Bplus} and \nr{Bminus}. 
These effects are present even in the massless
limit when helicity-conserving 
(fermion-number violating)
contributions are absent. 
In the massless limit the 
total lepton asymmetry equals minus the total helicity asymmetry 
integrated over momenta, cf.\ \se\ref{ss:fermion}.

As a final comment, 
we note that in a recent paper~\cite{recent}
a coupled set of evolution equations was derived,
within linear response theory, for the spin-averaged 
phase space distribution of 
one sterile neutrino species and for the total lepton asymmetry. 
Conceptually, this situation can be obtained from our framework by 
making two of the sterile neutrinos heavy
so that they represent ``fast variables''; 
integrating them out; averaging
over the helicity components of the light sterile neutrino; 
and restricting to leading order in deviations from
equilibrium. In practice
we cannot proceed to that limit because different approximations are
needed for treating fast and slow variables. Nevertheless, it would 
be interesting to understand whether analogues of the (small) 
``non-factorizable'' 
contributions of $\rmO(h^4)$ that were found in ref.~\cite{recent} 
could originate in our system, if our derivation were extended 
up to the $\rmO(h^4)$ level.  

%
\section*{Acknowledgements}

The role of helicity in the context of leptogenesis
has been considered by many people. 
J.G.\ thanks J.~L\'opez-Pav\'on for clarifications 
concerning ref.~\cite{n3} and for discussions.
M.L.\ thanks D.~B\"odeker for communicating results from a study by himself 
and D.~Schr\"oder, P.~Hern\'andez for discussions, 
and M.~Shaposhnikov for many exchanges
over the years. In particular, 
once the current manuscript had been finalized,  
we thank M.~Shaposhnikov for sharing with us a draft
by himself and S.~Eijima, in which helicity-flipping and conserving
rates similar to ours are discussed at $T < 130$~GeV.
This work was supported by the Swiss National Science Foundation
(SNF) under grant 200020-168988.

%
\appendix
\renewcommand{\thesection}{Appendix~\Alph{section}}
\renewcommand{\thesubsection}{\Alph{section}.\arabic{subsection}}
\renewcommand{\theequation}{\Alph{section}.\arabic{equation}}

%
\section{Origin of thermal mass corrections}

In this appendix we complement the derivation of 
\se\ref{se:derivation}, which concentrated on ``absorptive'' effects
(i.e.\ real scattering rates), by showing how the 
``dispersive'' thermal mass correction of \eq\nr{H0}
emerges within the same formalism (appendix~\ref{ss:mass}). 
We also take the opportunity
to display some steps of the general formalism in more detail, by
rederiving the main correlators 
within a quantum mechanical (bosonic) toy model
(appendix~\ref{ss:qm}). 

%
\subsection{Evolution equations in quantum mechanics}
\la{ss:qm}

Consider the quantum mechanical Hamiltonian 
\be 
 H = \underbrace{ \sum_k \Bigl( \omega^{ }_k a^\dagger_k a^{ }_k}_{H^{ }_0} 
 + h^*_k\, j^\dagger a^{ }_k + h^{ }_k a^\dagger_k\, j \Bigr)
 \; + \; 
 \mbox{(terms without $a,a^\dagger$)}
 \;, \la{toy_H} 
\ee
where $a^{ }_k$ and $a^\dagger_k$ are annihilation and creation
operators, and $j, j^\dagger$ are currents with which they interact. 
In the interaction picture (denoted with the subscript $I$), 
$a^{ }_k$ and $a^\dagger_k$ evolve with 
time: 
\be
 i \dot{a}^{ }_{kI}(t) = [a^{ }_{kI}(t),H^{ }_0] = \omega^{ }_k a^{ }_{kI}(t)
 \; \Rightarrow \; 
 a^{ }_{kI}(t) = a^{ }_k e^{-i\omega^{ }_k t}
 \;, \quad 
 a^{\dagger}_{kI}(t) = a^{\dagger}_k e^{i\omega^{ }_k t}
 \;. \la{def_a}
\ee
The time dependences here correspond to those in \eq\nr{onshell}. 
Like in the discussion below \eq\nr{ja}, we now go over to the Heisenberg
picture (denoted with the subscript $H$). Then
\ba
 i \partial^{ }_t (a^\dagger_{iH} a^{ }_{jH})
 & = & 
 [a^\dagger_{iH} a^{ }_{jH}, H]
 \; = \; 
 a^\dagger_{iH} [a^{ }_{jH}, H] + 
 [a^\dagger_{iH} , H] a^{ }_{jH}
 \nn 
 & = & 
 (\omega^{ }_j - \omega^{ }_i) a^\dagger_{iH} a^{ }_{jH}
 - h_i^* j^\dagger_H a^{ }_{jH} + h^{ }_j a^\dagger_{iH} j^{ }_H
 \;. \la{eom_1}
\ea
A density matrix associated with the particles
created by $a^\dagger$ is defined as 
\be
 \hat{\rho}^{ }_{ij} \; \equiv \; e^{i (\omega^{ }_j - \omega^{ }_i)t}
 a^\dagger_{iH} a^{ }_{jH} 
 \;. \la{qm_rho}
\ee
We note in passing that 
to $\rmO(h^0)$, when the Heisenberg and interaction pictures display
the same time evolution, we can identify
$
  \hat{\rho}^{ }_{ij} = a^\dagger_i a^{ }_j
$,
with the explicit time dependence in \eq\nr{qm_rho} cancelling
against that in \eq\nr{def_a}. 
Inserting \eq\nr{qm_rho} into \nr{eom_1} we get
\be
 i \dot{\hat{\rho}}^{ }_{ij} = 
 e^{i(\omega^{ }_j - \omega^{ }_i)t}
 \Bigl( 
 - h_i^* j^\dagger_H a^{ }_{jH} + h^{ }_j a^\dagger_{iH} j^{ }_H
 \Bigr)
 \;. 
\ee
The goal now is to evaluate the average of this operator in an ensemble
characterized by a density matrix $\rho^{ }_\rmi{full}$. 
For this task it is helpful to switch back into the interaction picture: 
\be
 \bigl\langle  i \dot{\hat{\rho}}^{ }_{ij}  \bigr\rangle
 \; \equiv \; 
 \tr \bigl[ i \dot{\hat{\rho}}^{ }_{ij}\, \rho^{ }_{\rmi{full}H} \bigr]
 \; = \; 
 \tr \Bigl[ \underbrace{\Bigr( 
  - h_i^*  j^\dagger_I  a^{ }_{j} e^{-i\omega^{ }_i t}
  + h^{ }_j  a^\dagger_{i} j^{ }_I e^{i\omega^{ }_j t}
 \Bigr)}_{i \dot{\hat{\rho}}^{ }_{ij I}} 
 \, \rho^{ }_{\rmi{full}I} \bigr]
 \;. \la{detailed}
\ee
Here we inserted interaction picture operators according 
to \eq\nr{def_a}. The time evolution of the interaction picture 
density matrix follows from \eq\nr{density_matrix}. Dropping the 
leading term because of odd discrete symmetries, and noting that two 
commutators do not contribute 
as explained below \eq\nr{gen_1}, we obtain
\ba
 \bigl\langle  \dot{\hat{\rho}}^{ }_{ij}  \bigr\rangle
 \; = \; 
 \int_0^t \! {\rm d}t' \, 
 \!\!&\!\! \displaystyle \sum_k \!\!&\!\! 
 \Bigl\langle
  h_i^* h^{ }_k e^{i(\omega^{ }_k t' - \omega^{ }_i t)}
  \bigl[ j^\dagger_{I}(t) a^{ }_j , a_k^\dagger j^{ }_I(t') \bigr]
 \nn & - & 
  h^{ }_j h^*_k e^{i(\omega^{ }_j t - \omega^{ }_k t')}
 \bigl[
  a_i^\dagger j^{ }_{I}(t),j^\dagger_{I}(t') a^{ }_k 
 \bigr]
 \Bigr\rangle + \rmO(h^3)
 \;.
\ea
The commutators can be simplified by making use of 
$
 [a^{ }_j,a_k^\dagger] = \delta^{ }_{jk}
$: 
\ba
  \bigl[ j^\dagger_{I}(t) a^{ }_j , a_k^\dagger j^{ }_I(t') \bigr]
 & = & 
 \delta^{ }_{jk}\, j_I^\dagger(t) j^{ }_I(t') + 
 a^\dagger_k a^{ }_j \bigl[ j^\dagger_I(t),j^{ }_I(t') \bigr]
 \;, \\ 
 \bigl[
  a_i^\dagger j^{ }_{I}(t),j^\dagger_{I}(t') a^{ }_k 
 \bigr]
 & = & 
 - \delta^{ }_{ik}\, j^\dagger_{I}(t') j^{ }_I(t) 
 + a^\dagger_i a^{ }_k
 \bigl[ j^{ }_I(t),j^\dagger_I(t') \bigr]
 \;. 
\ea
As explained just below \eq\nr{qm_rho}, the operators in the 
latter terms can be identified as $\hat{\rho}^{ }_{kj}$ and 
$\hat{\rho}^{ }_{ik}$, respectively, up to corrections 
of $\rmO(h)$. The ensemble averages of 
$j$ and $j^\dagger$ can be identified as advanced, retarded, and
Wightman correlators: 
\ba
 - \theta(t-t') \bigl\langle \bigl[ 
  j^\dagger_I(t),j^{ }_I(t')
  \bigr]\bigr\rangle & = & i \Pi^{ }_A(t'-t)
 \;, \\ 
 \theta(t-t') \bigl\langle \bigl[ 
  j^\dagger_I(t'),j^{ }_I(t)
  \bigr]\bigr\rangle & = & i \Pi^{ }_R(t-t')
 \;, \\
 \bigl\langle 
  j^\dagger_I(t) j^{ }_I(t')  
 \bigr\rangle & = & \Pi^{ }_{<}(t'-t)
 \;, 
\ea
where we made use of time-translation invariance. Thereby
\ba
 \bigl\langle  \dot{\hat{\rho}}^{ }_{ij}  \bigr\rangle
 & = & 
   h_i^* h^{ }_j e^{i(\omega^{ }_j - \omega^{ }_i)t}
  \int_0^t \! {\rm d}t' \, 
  \Bigl[
   e^{i \omega^{ }_j(t'-t)} \Pi^{ }_{<}(t'-t)
  + e^{i\omega^{ }_i(t-t')} \Pi^{ }_{<}(t-t') 
  \Bigr]
 \nn[2mm] 
 & + & \sum_k  
 \hat{\rho}^{ }_{ik} h_k^* h^{ }_j e^{i(\omega^{ }_j - \omega^{ }_k)t}
  \int_0^t \! {\rm d}t' \, 
  e^{i \omega^{ }_k(t-t')} i \Pi^{ }_R(t-t')
 \nn 
 & - &  \sum_k
 \hat{\rho}^{ }_{kj} h_i^* h^{ }_k e^{i(\omega^{ }_k - \omega^{ }_i)t}
  \int_0^t \! {\rm d}t' \, 
  e^{i \omega^{ }_k(t'-t)} i \Pi^{ }_A(t'-t)
  + \rmO(h^3)
 \;. \la{evol_toy} 
\ea
At this point we approximate 
$\omega^{ }_i \approx \omega^{ }_j \equiv \omega \gg 
|\omega^{ }_i - \omega^{ }_j|$
within the Fourier transforms, whereby
\ba
 \lim_{t\to\infty}
  \int_0^t \! {\rm d}t' \, 
  \Bigl[
   e^{i \omega^{ }(t'-t)} \Pi^{ }_{<}(t'-t)
  + e^{i\omega^{ }(t-t')} \Pi^{ }_{<}(t-t') 
  \Bigr] & = & \Pi^{ }_{<}(\omega) 
 \;, \la{Wight} \\ 
 \lim_{t\to\infty}
  \int_0^t \! {\rm d}t' \, 
  e^{i \omega^{ }(t-t')} i \Pi^{ }_R(t-t')
 & = & i \Pi^{ }_R(\omega)
 \;, \\ 
 \lim_{t\to\infty}
 \int_0^t \! {\rm d}t' \, 
  e^{i \omega^{ }(t'-t)} i \Pi^{ }_A(t'-t)
 & = & i \Pi^{ }_A(\omega)
 \;. 
\ea
The function $\Pi^{ }_{<} = 2 \nB{} \rho^{ }_\omega$ is real, 
whereas $i\Pi^{ }_{R,A}$ have both a real
and an imaginary part: 
$i\Pi^{ }_{R} = i \re \Pi^{ }_{R} - \rho^{ }_\omega$, 
$i\Pi^{ }_{A} = i \re \Pi^{ }_{R} + \rho^{ }_\omega$.
The real parts (proportional to the spectral function, 
denoted here by $\rho^{ }_\omega$ in distinction to the
density matrix $\rho$)
yield the absorptive effects
discussed in the main text.  
Focussing now on the dispersive imaginary parts and 
carrying out a substitution like 
in \eq\nr{time-dep}, we obtain a time evolution of the form 
$
 \dot{\rho}|^{ }_\rmi{dispersive} = i [M,\rho]
$, 
like in \eqs\nr{summary_rho_plus} and \nr{summary_rho_minus}, 
where
\be
 M^{ }_{ij} = -h_i^* h^{ }_j \re\Pi^{ }_R(\omega)
 \;. 
\ee
This matrix represents the ``standard'' energy correction 
for the system of \eq\nr{toy_H}. Its generalization to the case
of a Majorana fermion emerges through the first term in the dispersion 
relation in \eq\nr{spectral} and ultimately leads to \eq\nr{H0_IJ}. 

%
\subsection{Modified dispersion relation for ultrarelativistic sterile neutrinos}
\la{ss:mass}

Returning to the full system, 
consider the structure leading to the last term on 
the first row of \eq\nr{evol_rho} as an example.\footnote{%
 Thanks to its diagonal structure 
 the first term on the first row,
 containing the Fermi distribution, cancels against
 a contribution from the 
 corresponding term on the second row, once we work up to leading 
 order in the ultrarelativistic 
 approximation $\omega^k_\I \approx \omega^k_\J$.
 This is the same phenomenon which rendered the first row 
 of \eq\nr{evol_toy} into the purely real $\Pi^{ }_{<}$
 of \eq\nr{Wight}.  
 } 
Before restricting to the absorptive
part, this term reads 
\ba
 \left. \langle \dot{\hat{\rho}}^{ }_{\tau\I;\sigma\J } \rangle
 \right|_\rmi{first}
 & = & 
 - \sum_{\sL,a}
  e^{i (\omega^k_\sL - \omega^k_\I) t}
 \int_0^t \! {\rm d}t' 
 \int_{-\infty}^{\infty} \! \frac{{\rm d}\omega}{2\pi}
  e^{i(\omega - \omega^k_\sL)(t-t')}
 \nn 
 & \times &  
  \frac{h^*_{\I a}h^{ }_{\sL a}}{\sqrt{\omega^k_\I \omega^k_\sL}} \,
  \bar{u}^{ }_{\vec{k}\tau\sL} \, \aL \, \rho^{ }_a(\mathcal{K}^{ }) \, \aR \, 
  u^{ }_{\vec{k}\tau\I} \,
 \langle \hat{\rho}^{ }_{\tau\sL;\sigma\J} \rangle
 \;. \la{ex_first}
\ea
The integral over $t'$ can be carried out by making use of \eq\nr{principal}. 
Subsequently, we are faced with a spectral representation which can be 
identified as the real and imaginary parts of the retarded correlator: 
\be
 \int_{-\infty}^{\infty} \! \frac{{\rm d}\omega}{2\pi}
 \frac{i \rho^{ }_a(\omega,k)}{\omega - \omega^k_\sL + i \epsilon}
 \; = \; 
 \frac{i \re \Pi^{ }_\rmii{R}(\omega^k_\sL,k)
  + \rho^{ }_a (\omega^k_\sL,k) }{2}
 \;. \la{spectral}
\ee
The latter term leads to the absorptive behaviour in \eq\nr{evol_rho}, 
and we now focus on the first term. The retarded correlator is an 
analytic continuation of the Euclidean correlator, which for the 
operators in \eq\nr{ja} reads
\ba
 \Pi^{ }_\rmii{E}(K) & = & 
 \int_X e^{i K\cdot X}
 \bigl\langle j^{ }_a(X) \bar{j}^{ }_a(0) 
 \bigr\rangle \quad (\mbox{no sum over $a$})
 \nn 
 & = & 
 -2\, \Tint{\{P\}}
 \aL\, \frac{i \bsl{P}}{P^2 [(P+K)^2 + m_\phi^2]}\, \aR
 \;. 
\ea
At finite temperature 
the sum-integral is proportional to two independent Lorentz-tensors, 
$\bsl{K}$ and $\gamma^{ }_0$. 
After the analytic continuation $k^{ }_n \to -i (\ko + i 0^+)$, 
with $K=(k^{ }_n,\vec{k})$ and $\mathcal{K} = (\ko,\vec{k})$,  
so that 
$
 \Pi^{ }_\rmii{E}(K) \to 
 \Pi^{ }_\rmii{R}(\mathcal{K})
$, 
we can write 
\be
 \Pi^{ }_\rmii{R}(\mathcal{K}) = 
 \alpha\, \bsl{\mathcal{K}} + \beta \msl{u}
 \;, 
\ee
where $u = (1,\vec{0})$ is the four-velocity of the heat bath. 
After bracketing with on-shell spinors
according to \eq\nr{ex_first}, we are led to results 
similar to those in \eq\nr{rho_LPM}, specifically
\be
 \bar{u}^{ }_{\vec{k}\tau\sL}
  \, \aL \, 
  \bigl(   \alpha\, \bsl{\mathcal{K}}^{ }_{\sL} + \beta \msl{u} 
  \bigr) 
  \, \aR \, 
  u^{ }_{\vec{k}\tau\I}
 \; \approx \; 
 \left\{ 
  \begin{array}{ll}
  \displaystyle
   \alpha M^{ }_{\I} M^{ }_{\sL}
  + \frac{ \beta M^{ }_{\I} M^{ }_{\sL}}{2 k}\;, & \tau = - \\ 
  \displaystyle
   \alpha M_{\sL}^2 + 
   \beta \,\bigl( 2 k + \frac{M^{2}_{\I}
  +  M^{2}_{\sL} }{4k} \bigr)\;, & \tau = + 
  \end{array}
 \right.  
 \;. \la{trace2}
\ee
For the opposite chiral projections, the roles of the helicity
states are exchanged. In any case, 
for $k \gg M^{ }_{\sL}$, only the contribution 
proportional to $\beta$ is needed. 

We can write $\beta = 2 \mathcal{V}(m^{ }_\phi)/k$, 
where $\mathcal{V}$ is given in \eq(5.10) of ref.~\cite{broken}.
In particular, for $\pi T \gg m^{ }_\phi$ we get
$\beta \approx -T^2 / (8 k)$. Recalling the factors from 
\eqs\nr{spectral} and \nr{trace2}, this yields
\be
 \left. \langle \dot{\hat{\rho}}^{ }_{\tau\I;\sigma\J } \rangle
 \right|_\rmi{first} = 
  \sum_{\sL,a} \frac{i h^*_{\I a}h^{ }_{\sL a} T^2
  \, \delta^{ }_{\tau,+}}{8 k}
 \langle \hat{\rho}^{ }_{\tau\sL;\sigma\J} \rangle
 \;. 
\ee
Adding the three other channels and going over to the notation of 
\eq\nr{def_rho_new} produces
\be
 \left. \dot{\rho}^{ }_{(\tau)} \right|^{ }_\rmi{dispersive} 
 \approx 
 i [H^{ }_{0(\tau)},\rho^{ }_{(\tau)}]
 \;,
\ee
where 
\be 
 H^{ }_{0(\tau)\I\J} = 
 \delta^{ }_{\I\J} \, \omega^k_\I + 
 \sum_a 
 \frac{(
     h^{ }_{\I a}h^{*}_{\J a} \, \delta^{ }_{\tau,-}
   + h^*_{\I a}h^{ }_{\J a} \, \delta^{ }_{\tau,+}
  ) T^2
  }{8 k}
 \;. \la{H0_IJ}
\ee
After symmetrizing or antisymmetrizing in helicity, 
this leads to \eqs\nr{H0} and \nr{Delta0}. 

%

\end{document}